# Phonon and Photon Lasing Dynamics in Optomechanical Cavities

Jian Xiong[1,2,†], Zhilei Huang[1,2,†], Kaiyu Cui [1,2,*], Xue Feng[1,2], Fang Liu[1,2], Wei Zhang[1,2,3], and Yidong Huang[1,2,3*]

[†]These authors contributed equally to this work

* Corresponding author: <u>kaiyucui@tsinghua.edu.cn; yidonghuang@tsinghua.edu.cn</u>

[1]Department of Electronic Engineering, Tsinghua University, Beijing 100084, China

[2]Beijing National Research Center for Information Science and Technology, Beijing, China

[3]Beijing Academy of Quantum Information Science, Beijing, China

## Abstract

Lasers differ from other light sources in that they are coherent, and their coherence makes them indispensable to both fundamental research and practical application. In optomechanical cavities, phonon and photon lasing is facilitated by the ability of photons and phonons to interact intensively and excite one another coherently. The lasing linewidths of both phonons and photons are critical for practical application. However, thus far, these linewidths have not been explored in detail in cavity optomechanical systems. This study investigates the underlying dynamics of lasing in optomechanical cavities and experimentally demonstrates simultaneous photon and phonon lasing with narrow linewidths in a silicon optomechanical crystal cavity. We find that the linewidths can be accounted for by two distinct physical mechanisms in two regimes, namely the normal regime and the reversed regime, where the intrinsic optical decay rate is either larger or smaller than the intrinsic mechanical decay rate. In the normal regime, an ultra-narrow spectral linewidth of 5.4 kHz for phonon lasing at 6.22 GHz can be achieved regardless of the linewidth of the pump light, while these results are counterintuitively unattainable for photon lasing in the reversed regime. These results pave the way towards harnessing the coherence of both photons and phonons in silicon photonic devices and reshaping their spectra, potentially opening up new technologies in sensing, metrology, spectroscopy, and signal processing, as well as in applications requiring sources that offer an ultra-high degree of coherence.

As the strongest and most tunable nonlinear interaction, the photon-phonon interaction can be harnessed in both classical and quantum regimes[1]. Brillouin scattering, which results from interactions between photons and travelling acoustic phonons, has shown considerable potential for high-coherence and ultra-low-noise oscillation owing to the unique lasing dynamics of stimulated Brillouin scattering (SBS)[2,3]. SBS in ordinary optical materials such as glass has facilitated a range of technological applications in metrology[4], sensing[5,6], and signal processing[7,8]. However, achieving SBS in silicon, the fundamental material underlying photonic integrated circuits, is challenging because the acoustic wave in a silicon waveguide irreversibly radiates into the silica cladding layer. Specifically, Brillouin lasing[9] and modulation[10] are possible only through the use of silicon waveguides that are several centimetres long and suspended in air[11,12]. It is technically challenging to build such structures, which are important for laser technologies based on silicon[9–12].

Photon-phonon interactions can be greatly enhanced by optomechanical cavities[13,14]. Such cavities have facilitated the demonstration of many significant phenomena[15–18]. For example, phonon lasers, which are mechanical analogues of photon lasers, have attracted considerable attention owing to their underlying intriguing physics[19–22] and unprecedented capability for ultra-precise sensing[23]. The linewidth limit of phonon lasers is theoretically predicted to follow a Schawlow–Townes form, similar to that of photon lasers[24]. In addition, optomechanical limit cycle theory[25] has been developed as an efficient tool for studying the physics of lasing dynamics, providing analytical solutions of the phonon linewidth with thermal noise in the normal regime[26]. For cavity optomechanical systems, this theory has successfully explained the dynamical multistability phenomenon[27], amplitude noise suppression[28], and phase noise using linear approximation of the Langevin equation[26]. Despite these significant theoretical advances, the phase noise of the pump light, which severely limits the realisation of ultra-narrow linewidth for photon and phonon lasers in practice, has not been considered in theoretical studies thus far.

Phase noise, which varies inversely with oscillator energy in accordance with the classical theory, has been experimentally observed with phonon lasers[29,30]. Other experimental studies have shown that, after phonon lasing, the photons scattered by phonons exhibit coherent light conversion[31]. Such coherent conversion is not only important in the quantum regime[15] but also indicates that optomechanical cavities may offer a solution for reshaping the spectra and coherence of both photons and phonons, thereby facilitating new applications in sensing, metrology, spectroscopy, signal processing, and other fields that require highly coherent sources. According to previous studies[32,33], photons and phonons do not lase simultaneously: mechanical self-oscillation occurs with the narrowing of the mechanical line in the normal regime, while optical self-oscillation occurs with the narrowing of the optical line in the reversed regime. The possibility of excitation and simultaneous lasing of photons and phonons in optomechanical cavities has not been identified thus far. Such identification is necessary for wider applications of coherent conversion.

Coherent photon-phonon interactions may also complement silicon-based Brillouin lasers[9,34]. Optical self-oscillation and true line narrowing to the Schawlow–Townes limit have been reported to require the reversed dissipation regime in Brillouin lasers[33]. However, in contrast to a previous study[32], we find that high-coherence and ultra-low-noise oscillation for the scattered light in optomechanical cavities is counter-intuitively unattainable in the reversed regime. Therefore, it is important to investigate the underlying lasing dynamics in optomechanical cavities because the phase noise of the coherent photons and phonons, observed in the form of linewidth broadening, is critical to a wide range of practical applications.

This study is the first to reveal the simultaneous lasing of photons and phonons as well as their linewidths in an cavity optomechanical system. The lasing dynamics considering both the pump noise and the thermal noise are explored by a general limit cycle theory without linearisation approximation. Thus, the lasing phenomena can be explained in both normal and reversed regimes. The general expressions of the analytical solutions for the lasing linewidth are obtained, which are in good agreement with the numerical analysis. The linewidth-narrowing mechanism can be explained on the basis of a phase-noise-following scenario, and ultra-narrow spectral linewidth for phonon lasing can be realised regardless of the linewidth of the pump light. However, such results are counterintuitively unattainable for photon lasing, which differs significantly from a Brillouin system. We experimentally demonstrate simultaneous photon and phonon lasing in a silicon-based optomechanical crystal (OMC) cavity with an acoustic radiation shield. We observe a phonon lasing frequency of up to 6.22 GHz, with the scattering photon lasing at a communication-relevant wavelength of 1533.9 nm. The optomechanical coupling rate of the cavity is 1.9 MHz. The observed linewidth of the excited phonon decreases to 5.4 kHz, which is four orders of magnitude narrower than the linewidth of the pump light, whereas the linewidths of the scattering photon and pump light are similar, which is in agreement with our theoretical predictions. Harnessing the coherence of photons and phonons in silicon devices could pave the way towards intriguing new applications in sensing, metrology, spectroscopy, and signal processing.

## Results

### Lasing dynamics of phonon and scattering photon

In an cavity optomechanical system, the light interacts with the mechanical motion, which affects the effective mechanical damping rate. When we blue-detune the pump light and increase its power, the effective mechanical damping can be not only reduced to zero but also changed to amplification. Accordingly, coherently self-sustained oscillation, i.e. phonon lasing, and stimulated photon scattering can occur with coherent light conversion. The combination of phonon lasing and stimulated photon scattering is shown in Fig. 1. Note that simultaneous phonon and photon lasing can occur in optomechanical cavities because phonons and photons show identical threshold values as shown in Figure 4a. The experimental data in this figure indicate that large energy accumulation of phonons and photons will occur simultaneously if the pump power exceeds the laser threshold (the details are provided in Supplementary Information section S3-2).

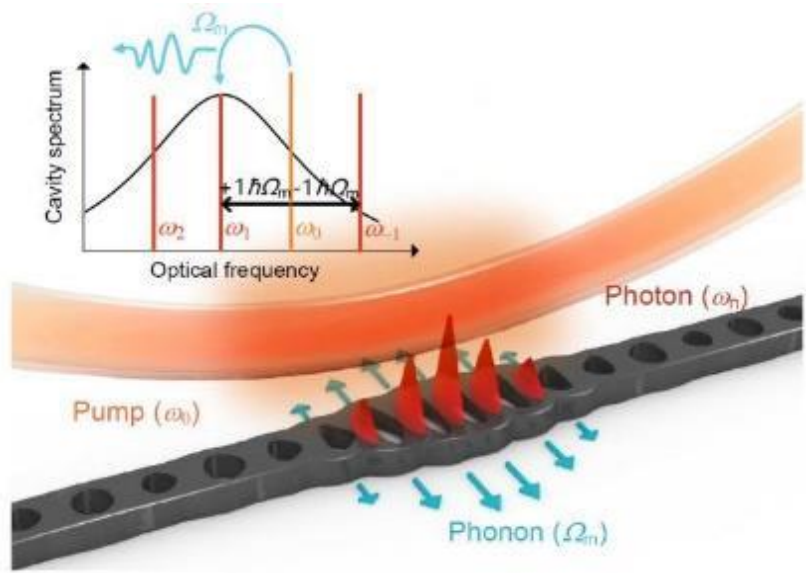

**Figure 1.** Schematic of simultaneous photon/phonon lasing. The pump photons (shown in orange, $\omega_0$), which are coupled to the cavity with a tapered fibre, are transferred to the scattering photons (shown in red, including the high-order scattering light), and they excite the phonons (shown in blue). The photon with frequency $\omega_n$ represents the scattering photon converted from the pump photon by releasing $n$ phonons.

For the cavity optomechanical system, the interaction Hamiltonian between a photon and a phonon can be written as $-\hbar g_0 \hat{a}^\dagger \hat{a}(\hat{b}^\dagger + \hat{b})$, where $\hbar$ is the reduced Planck constant, $g_0$ is the vacuum optomechanical coupling strength, and $\hat{a}\,(\hat{a}^\dagger)$ and $\hat{b}\,(\hat{b}^\dagger)$ are the annihilation (creation) operators of photons and phonons, respectively[35]. Based on this interaction Hamiltonian, the evolution equations of the cavity system can be derived in the semi-classical regime as

$$\dot{a} = (i\Delta - \kappa/2)a + ig_0(b^* + b)a + \sqrt{\kappa_{ex}}\, s_{in}, \tag{1}$$

$$\dot{b} = -(i\Omega_m + \Gamma/2)b + ig_0|a|^2 + \sqrt{\Gamma}\, b_{th}, \tag{2}$$

where $a$ is the optical amplitude in the cavity; $\Delta$ represents the laser detuning from the cavity resonance, i.e. $\omega_0 - \omega_1$, where $\omega_1$ is the optical angular frequency of the cavity and $\omega_0$ is the angular frequency with which the frame rotates with the pump laser; $\kappa$ is the total decay rate of the optical mode; $\kappa_{ex}$ is the optical coupling rate between the optical input channel and the cavity; $b$ is the mechanical amplitude that satisfies $x = x_{zpf}(b+b^*)$, where $x$ is the displacement of the mechanical cavity and $x_{zpf}$ denotes the zero-point fluctuation displacement; $\Omega_m$ represents the intrinsic angular frequency of the mechanical mode; $\Gamma$ is the decay rate of the mechanical mode; $b_{th}$ is the noise of the mechanical oscillator owing to the thermal fluctuations in the environment, which is expressed as $b_{th} = \sqrt{n_{th}}\dot{W} = \sqrt{k_B T/\hbar\Omega_m}\dot{W}$, where $n_{th}$ is the number of phonons in the thermal equilibrium state $k_B$ is the Boltzmann constant, $T$ denotes temperature, and $W$ is the standard Wiener process[36]; and $s_{in}$ is the amplitude of the pump light[37,38]. As the intensity noise can be ignored for a typical single-mode laser, the linewidth of the pump laser originates from the phase noise[34,39]. Thus $s_{in} = |s_{in}|e^{i\varepsilon_p}$, where $\varepsilon_p = \sqrt{v_p}W$ and $v_p = \text{var}(\dot{\varepsilon}_p)$ is the angular frequency linewidth of the pump laser[40].

According to the limit cycle theory, the scattering photons and phonons in the cavity can be written in the form of a Fourier series expansion, i.e. $b = B_c + Be^{-i\Omega_m t}$ and $a = \sum_n \alpha_n e^{in\cdot\Omega_m t}$, where $B_c$ represents the equilibrium position of the mechanical vibration and $B$ is the normalised complex amplitude of the mechanical vibration (including the amplitude $B_0$ and the phase $\phi$, i.e. $b = B_c + B_0 e^{i\phi}e^{-i\Omega_m t}$). Here, we consider that the mechanical vibration mode with multiple frequencies of $\Omega_m$ usually has a large dissipation rate such that only the first harmonic term is retained[26]. Further, $\alpha_n$ is the normalised complex amplitude of photons with different frequencies $\omega_n$ in the cavity, i.e. $\alpha_n = A_n e^{i\varepsilon_n}$ ($A_n$ is the amplitude of the optical field and $\varepsilon_n$ is the corresponding phase). The subscript $n$ implies that the scattering photon with index $n$ comes from the pump photon after releasing $n$ phonons. For instance, $\alpha_0$ ($A_0 e^{i\varepsilon_0}$) represents the normalised complex amplitude of the photon with frequency $\omega_0$. By applying Fourier series expansion of $a$ and $b$ to Eqs. (1) and (2), the equations for analysing the phase dynamics can be obtained (see Supplementary Information section S3).

$$\dot{\phi} = \frac{M\Gamma}{2}\cos(\phi+\varepsilon_1-\varepsilon_0) + \frac{\sqrt{\Gamma}}{\langle B_0 \rangle}b_{th}\sin(\Omega_m t - \phi) \qquad (3)$$

$$\dot{\varepsilon}_1 = \tilde{\Delta} - \Omega_m + \frac{M\kappa}{2}\cos(\phi+\varepsilon_1-\varepsilon_0) \qquad (4)$$

$$\dot{\varepsilon}_0 = \tilde{\Delta} + g_0 \left\langle \frac{B_0 A_1}{A_0} \right\rangle \cos(\phi + \varepsilon_1 - \varepsilon_0) + \frac{\sqrt{\kappa_{ex}} |s_{in}|}{\langle A_0 \rangle} \sin(\varepsilon_p - \varepsilon_0) \tag{5}$$

The above-mentioned equations are the phase reduction equations, which are often used to analyse the phase evolution and synchronisation phenomena in the self-sustained oscillating system[41]. Here, $M$ is a nonlinear factor that characterises the strength of the pump and $\tilde{\Delta}$ represents the equivalent detuning. When the system is lasing, the phonon phase $\phi$, the scattering photon phase $\varepsilon_1$, and the pump light phase $\varepsilon_p$ need to be synchronised (see Supplementary Information section S3-2). In this paper, we introduce this phase synchronisation condition to solve Eqs. (3)–(5) (see Supplementary Information section S3 for details). Based on the phase reduction equations, the angular frequency linewidths of the phonon ($\nu_m$) and the scattering photon ($\nu_s$) can be obtained from the variance of the derivative of phase: $\nu_m = \mathrm{var}(\dot{\phi})$ and $\nu_s = \mathrm{var}(\dot{\varepsilon}_1)$. We need to consider only the random variables, namely $\phi$, $\varepsilon_0$, $\varepsilon_1$, $b_{th}$, and $\tilde{\Delta}$ on the right-hand side of Eqs. (3)–(5), as the constant terms do not affect the linewidth. First, we analyse the influence of $\tilde{\Delta}$. The fluctuation strength of $\tilde{\Delta}$ can be derived as follows (see Supplementary Information section S3-1 for details):

$$\tilde{\Delta} = \Delta + 2g_0 \mathrm{Re}[B_c] = \Delta + \chi^{(2)} \sum |a_n|^2 \tag{6}$$

$$\delta\tilde{\Delta} = \tilde{\Delta} - \langle\tilde{\Delta}\rangle = 2\chi^{(2)} \sum A_n \delta A_n \quad \text{(keep the first-order fluctuation term)}$$

$$= 2\chi^{(2)}(A_1 \delta A_1 + A_0 \delta A_0) \quad \text{(use the sideband resolved assumption)}$$

$$= 2\chi^{(2)}\left[(\Gamma/\kappa) B_0 \delta B_0 + A_0 \delta A_0\right] \quad \text{(use the energy conservation relation } A_1/B_0 = \sqrt{\Gamma/\kappa}) \tag{7}$$

It can be concluded from Eq. (6) that $\tilde{\Delta}$ comes from the random jitter of the equilibrium position $B_c$, which is determined by the radiation pressure. Hence, $\tilde{\Delta}$ is associated with the intra-cavity field strength $\sum |a_n|^2$ and is thus affected by the fluctuation of the intra-cavity field intensity. From the expression of $\tilde{\Delta}$, $\delta\tilde{\Delta}$ can be derived as Eq. (7). Here, $\chi^{(2)} = 2g_0^2/\Omega_m$ is the equivalent Kerr nonlinear coefficient[25], which indicates the influence of the field intensity on the effective cavity length. Further, $\delta A_n = A_n - \langle A_n \rangle$ represents the amplitude noise of the nth-order scattering photon. When there are fluctuations in the field intensity, the equivalent cavity length will undergo random jitter. As the amplitude noise in the limit cycle oscillator is suppressed[28], only the first-order fluctuation term needs to be retained.

It can be seen that the effective detuning $\delta\tilde{\Delta}$ is directly associated with the dissipation rate $\Gamma/\kappa$. For the normal regime ($\Gamma \ll \kappa$), the ratio of $\Gamma/\kappa$ is considerably small; hence, we can ignore the first term.

Moreover, the pump strength $A_0$ can also be ignored because a phonon with a sufficiently low dissipation rate requires only a low-power pump to achieve lasing. Note that in this regime, treating $\tilde{\Delta}$ as a constant is a common strategy[35]. Through the analysis presented above, we conclude that $\delta\tilde{\Delta}$ can be ignored in the normal regime. However, the situation is different for the reversed regime. In this case, we can still ignore $A_0$ because that there is only one dominant optical mode $\omega_1$ in this regime[32] (see Supplementary Information Fig. S2 for details). However, the first term in Eq. (7) cannot be ignored owing to the large ratio of $\Gamma/\kappa$. Hence, the fluctuation in the effective detuning $\delta\tilde{\Delta}$ will be remarkable in the reversed regime. As the effects of $\delta\tilde{\Delta}$ on the linewidth characteristics are different in the two regimes, we will discuss these two regimes separately. In addition, we will directly solve the temporal stochastic differential equation for Eqs. (1) and (2) numerically without any approximation to validate our theoretical results.

**Normal regime** ($\Gamma \ll \kappa$): As the mechanical decay rate is much smaller than the optical decay rate, the phonons can be accumulated more effectively compared with the photons. Thus, $\tilde{\Delta}$ can be treated as a constant, i.e. $\tilde{\Delta} \approx \Delta$, which can be derived from Eq. (6) as well when the jitter of the intra-cavity light field strength $\sum|a_n|^2$ is negligible. Under this condition, we obtain the linewidths of both the scattering photons and the phonons by solving the phase reduction equations (3)–(5) (see Supplementary Information section S3-3-1 for details):

$$v_s = v_{mT} + \frac{v_p}{(1+\Gamma/\kappa)^2} \tag{8}$$

$$v_m = v_{mT} + \frac{v_p}{(1+\kappa/\Gamma)^2} \tag{9}$$

$$v_{mT} = \frac{\Gamma n_{th}}{2n} \tag{10}$$

The linewidth expression above is divided into two parts. One part is the linewidth of the Schawlow–Townes (ST) type, $v_{mT}$, which is inversely proportional to the number of intra-cavity phonons $n$. When the ratio of $n$ to the number of thermal equilibrium phonons $n_{th}$ is large, the broadening of the linewidth owing to thermal noise will be strongly suppressed. The ST linewidth formula derived from our equations is consistent with that obtained by Kerry J. Vahala[24]. The other part is the linewidth broadening owing to the phase noise of the non-ideal pumping light source, which is similar to the linewidth in SBS[34]. Besides the analytical expressions for the lasing linewidth, in the normal regime ($\kappa \gg \Gamma$), we can also obtain the important relation $\dot{\varepsilon}_1 = \tilde{\Delta} - \Omega_m + (\kappa/\Gamma)\dot{\phi}$ (derived from Eqs. (3) and (4) by ignoring the thermal noise). Further, by assuming that $\tilde{\Delta} \approx \Omega_m$, the reduced expression $\dot{\varepsilon}_1 = (\kappa/\Gamma)\dot{\phi}$ can be obtained. This is a good theoretical support for the approximation of $\dot{\phi} \ll \dot{\varepsilon}_1$, which is widely used in the normal regime without any theoretical explanation[25,28,42].

We also directly solve the temporal stochastic differential equation for Eqs. (1) and (2) numerically, and we use the Fourier transform to obtain the spectrum in the frequency domain. The spectra are shown in Fig. 2.

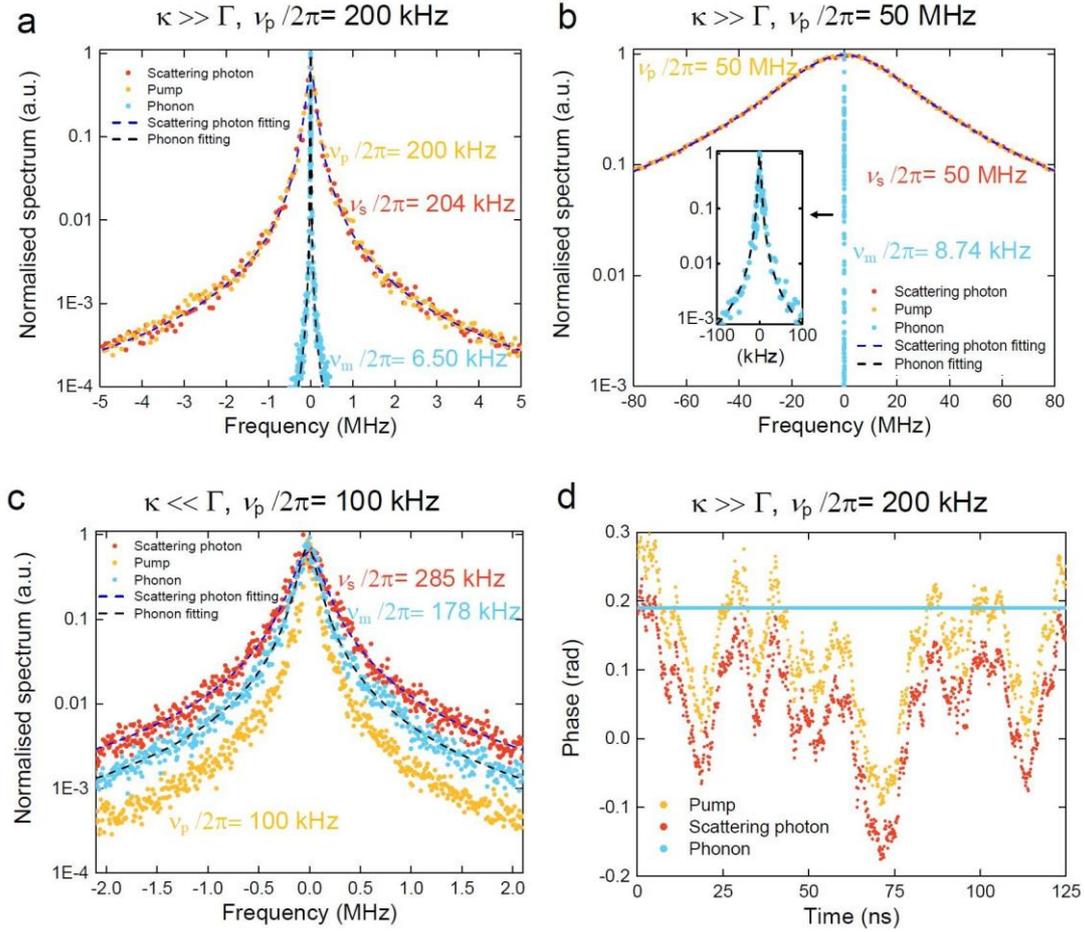

**Figure 2.** Simulated simultaneous photon and phonon lasing results. (**a**)–(**c**) Spectra of the pump, scattering photon, and phonon (mechanical motion) after lasing. The linewidth values marked in the figure are obtained from the Lorentz fitting. Further, we use the derived analysis expressions (8), (9), and (13) to calculate the theoretical linewidths. They are consistent with the Lorentz fitting. Results obtained with a cavity in the normal regime ($\kappa \gg \Gamma$) at a pump-laser linewidth of (**a**) 200 kHz and (**b**) 50 MHz. (**c**) Results of the cavity in the reversed regime ($\kappa \ll \Gamma$) with a pump-laser linewidth of 100 kHz. (**d**) Phase fluctuation of the pump, scattering photon, and phonon after lasing in the normal regime ($\kappa \gg \Gamma$) with a pump-laser linewidth of 200 kHz for the system represented in **a**.

For the normal regime, Fig. 2a shows the simulated spectrum of the cavity optomechanical system after lasing. The parameters were set as $\kappa/2\pi = 30$ GHz, $\Omega_m/2\pi = 6.2$ GHz, and $\Gamma/2\pi = 3.2$ MHz, which were the values in our experiments. The linewidth of the pump light ($v_p/2\pi$) was set to $2.0\times 10^2$ kHz. We can see that after lasing, the spectral linewidth of the scattering photon ($v_s/2\pi$) is around $2.0\times 10^2$ kHz, which is nearly the same as that of the pump light. By contrast, the spectral linewidth of the phonon ($v_m/2\pi$), which is around 6.5 kHz, is much narrower than that of the pump light. The spectral linewidths are also calculated on the basis of the analytical expressions of Eqs. (8) and (9). The calculated $v_s/2\pi$ and $v_m/2\pi$ are about 206.35 kHz and 6.35 kHz, respectively, which are in good agreement with the numerical analysis.

These results indicate that the phonon linewidth is several orders of magnitude narrower than that of the scattering photon for a cavity with $\kappa \gg \Gamma$ after lasing. To further understand the reason for this difference, the phase noise of the scattering photon and phonon was simulated (Fig. 2d). We find that the scattering photon follows the phase fluctuations of the pump light, while the phonon preserves its phase because the damping rate of the scattering photon is greater than that of the phonon. We can also use the adiabatic approximation to understand the transmission mechanism of the phase noise of the pump light. When the phonon dissipation rate is extremely low, the phonon is nearly adiabatic, which means that it needs nearly no assistance from the pump to maintain its stability. Hence, the phase noise of the pump light cannot be transmitted to the phonon, and the broadening of the phonon spectral line can thus be ignored, which is also indicated by the narrowing factor $(1+\kappa/\Gamma)^2 \to \infty$ in Eq. (9). However, the scattering photon has a large dissipation rate and thus succeeds the phase fluctuations of the pump light, which is shown by the narrowing factor $(1+\Gamma/\kappa)^2 \to 1$ in Eq. (8). When the optomechanical cavity is at room temperature, for the normal regime ($\Gamma \ll \kappa$), the thermal noise dominates the linewidth of the phonon with an ST-type formula[24].

Note that photon and phonon lasing can also be excited by a pump whose linewidth is much greater than the intrinsic mechanical linewidth. For instance, the parameter setting for the simulation shown in Fig. 2b is the same as that shown in Fig. 2a, except that $v_p/2\pi$ was changed to 50 MHz, which was much greater than the intrinsic mechanical linewidth of 3.3 MHz. As can be seen in Fig. 2b, mechanical oscillation can also be excited by this high-noise pump light with a linewidth of 8.74 kHz, because the ST linewidth is dominant in this regime and the change in the pump linewidth thus has a very small impact on the phonon. The slightly broadening of the phonon linewidth is owing to the decrease of the phonon number, which is caused by the increased linewidth of the pump. This ultra-narrow linewidth has considerable potential for applications in various fields such as high-precision sensing, high-quality frequency references, frequency-coherent conversion[43], and low-noise phonon sources[44].

**Reversed regime**($\Gamma \gg \kappa$): This spectral region is beyond the scope of existing silicon-based microstructures because the required optical Q-factor is rather high (typically, it needs to exceed $3.2 \times 10^8$ for a 6-GHz mechanical mode with a Q-factor of $1 \times 10^3$ and $\Gamma/\kappa = 10$). However, this scenario can be realised in the microwave domain[45]. Nevertheless, an important question is whether the scattering photon undergoes linewidth narrowing similar to the phenomenon in the case of a Brillouin laser regardless of the materials used or the design of the structures employed.

To answer this question, we must re-examine the linewidth of the scattering photon and phonon from Eqs. (3)–(5) by considering the influence of the equilibrium position. (a) The driving force (radiation pressure) term of the mechanical vibration is the beat note of the scattering light and pump light, which is related only to their relative phase difference $\varepsilon_1 - \varepsilon_0$. (b) The modulation due to the jitter in the equilibrium position is the same for every intra-cavity optical spectrum line. Therefore, the random phase of the phonon, which is determined only by the beat note of the optical line, will not be affected by such jitter and it still satisfies Eq. (9), which means that in the reversed regime ($\Gamma \gg \kappa$), $v_m \approx v_p/(1+\kappa/\Gamma)^2 + v_{mT}$. This conclusion can also be drawn from the interaction picture, i.e. the jitter in the equilibrium position can modulate the phase of the photons by the scattering process, but there is no direct interaction between the mechanical vibration and the cavity length. Hence, the phase of the phonons should not be affected. To further analyse the linewidth of the scattering photons, we define $\dot{\varepsilon}_c = \delta\tilde{\Delta}$ representing the random phase diffusion due to the jitter in the of equilibrium position such that the original phase of the photons before perturbation by the jitter can be separated, i.e. $\dot{\varepsilon}'_n = \dot{\varepsilon}_n - \dot{\varepsilon}_c$. In this manner, Eqs. (4) and (5) can be reduced to (see Supplementary Information section S3-3-2 for details)

$$\dot{\varepsilon}'_1 = \langle\tilde{\Delta}\rangle - \Omega_m + \frac{M\kappa}{2}\cos(\phi + \varepsilon_1 - \varepsilon_0) \tag{11}$$

$$\dot{\varepsilon}'_0 = \langle\tilde{\Delta}\rangle + g_0\left\langle\frac{B_0 A_1}{A_0}\right\rangle\cos(\phi + \varepsilon_1 - \varepsilon_0) + \frac{\sqrt{\kappa_{ex}}|s_{in}|}{\langle A_0\rangle}\sin\left[(\varepsilon_p - \varepsilon_c) - \varepsilon_0\right] \tag{12}$$

If $(\varepsilon_s - \varepsilon_c)$ is considered to be the pseudo pump phase, when we ignore $\delta\tilde{\Delta}$ in Eqs. (3)–(5), the form of Eqs. (3), (11), and (12) is similar to that of Eqs. (3)–(5). Thus, we can use the same method as that for the normal regime to obtain the linewidth formula. Combining the relation $v_m \approx v_{mT} + v_p/(1+\kappa/\Gamma)^2$ derived above, we obtain (see Supplementary Information section S3-3-2 for details)

$$v_s = v_{mT} + 2v_p + v_p/(1+\Gamma/\kappa)^2 \approx v_{mT} + 2v_p \tag{13}$$

$$v_m = v_{mT} + \frac{v_p}{(1+\kappa/\Gamma)^2} \quad \text{(phonon linewidth formula is the same in both regimes)}$$

$$\nu_{mT} = \frac{\Gamma n_{th}}{2n} \qquad \text{(ST linewidth formula is the same in both regimes)}$$

In the reversed regime ($\Gamma \gg \kappa$), owing to the strong jitter in the equilibrium position, the phase noise of the pump can be transmitted to the scattering photons, which overcomes the adiabatic approximation of the scattering photons. Thus, even if the intensity of the intra-cavity scattering photons is high when the system is lasing, the scattering photon linewidth $\nu_s$ cannot reach the ST limit, which implies that the self-purifying spectral line of the scattering photons, similar to that exhibited in SBS of the photon laser regime, is impossible to achieve in optomechanical cavities. Further, if we accidentally ignore the jitter in the equilibrium position (high-order terms of the interaction Hamiltonian), we can, by contrast, conclude that the spectral line can be narrowed to the ST limit[32]. This discrepancy is counterintuitive when compared with the Brillouin system, because there is no equilibrium position for the travelling acoustic wave in the Brillouin system.

The numerically simulated spectrum after lasing for an cavity optomechanical system with $\kappa \ll \Gamma$ in the reversed regime is shown in Fig. 2c. Here, we set $\kappa/2\pi = 6.5$ MHz, $\Gamma/2\pi = 197$ MHz, and $\nu_p/2\pi = 100$ kHz in the simulation. As shown in Fig. 2c, $\nu_s/2\pi$ and $\nu_m/2\pi$, with simulated values of about 285kHz and 178kHz, respectively, are both broader than the linewidth of the pump light. From Eqs. (9) and (13), the calculated linewidths $\nu_s/2\pi$ and $\nu_m/2\pi$ are 285.8 kHz and 185.8 kHz ($\nu_{mT}/2\pi$=85.8 kHz), which indicates that the theoretical analysis is well consistent with our simulation results.

To summarise the lasing dynamics, we find that the linewidth of photon and phonon lasing can be explained and analysed with reference to two distinct physical regimes ($\kappa \gg \Gamma$ and $\kappa \ll \Gamma$). For the normal regime ($\kappa \gg \Gamma$), the linewidth of the phonon is much narrower than that of the pump laser, and the ultra-low-noise mechanical signal can be excited even with a high-noise pump light whereas the linewidths of the scattering photon and pump light are similar. This linewidth-narrowing mechanism is similar to that of the Brillouin laser, except that the roles of the phonon and the scattering photon are reversed, as the typical Brillouin system satisfies the $\kappa \ll \Gamma$ condition. Meanwhile, for the reversed regime ($\kappa \ll \Gamma$), the linewidths of both the scattering photon and the phonon are broader than the linewidth of the pump light. The counterintuitive linewidth of the scattering photon originates from the jitter in the equilibrium position of the optomechanical cavity system and it differs significantly from that of the Brillouin system.

**OMC cavity with acoustic radiation shield**

An OMC cavity with an acoustic radiation shield was fabricated to demonstrate the photon and phonon lasers in practice. An oblique view of a scanning electron microscope (SEM) image of the structure is shown in Fig. 3a. The confinement mechanism of this structure is different from that of conventional OMC cavities, where the optical and mechanical modes are simultaneously confined by the same periodic structure[46]. As the mechanical frequency of this structure exceeds the phononic bandgap of nanobeam periodic structures, the nanobeam periodic structures of this OMC cavity can only confine the optical mode. By contrast, the acoustic radiation shield shown in Fig.3b confines the mechanical mode as its phononic bandgap covers the mechanical frequency (the details of the band structures are provided in Supplementary Information section S9). Thus, the optical and mechanical modes are separately confined by two types of structures. This confinement mechanism is used to overcome the design constraints of the optical and mechanical properties[47]. Owing to the design flexibility offered by this separate confinement mechanism, the mechanical frequency of the cavity reaches 6.22 GHz while the optical mode is maintained at 1533.9 nm. The intrinsic optical Q-factor and mechanical Q-factor in an ambient environment are $8.5\times10^3$ and $2.0\times10^3$, respectively, which are indicated by the optical transmission spectrum and the mechanical spectrum as shown in Fig. 3c and d.

In addition, owing to the design flexibility offered by the separate confinement of the optical and mechanical modes in the hetero structure, we were able to demonstrate a strong optomechanical coupling. The measured optomechanical coupling rate ($g_0/2\pi$) of the cavity was 1.9 MHz, which is the highest among reported optomechanical systems (see Supplementary Information section S6).

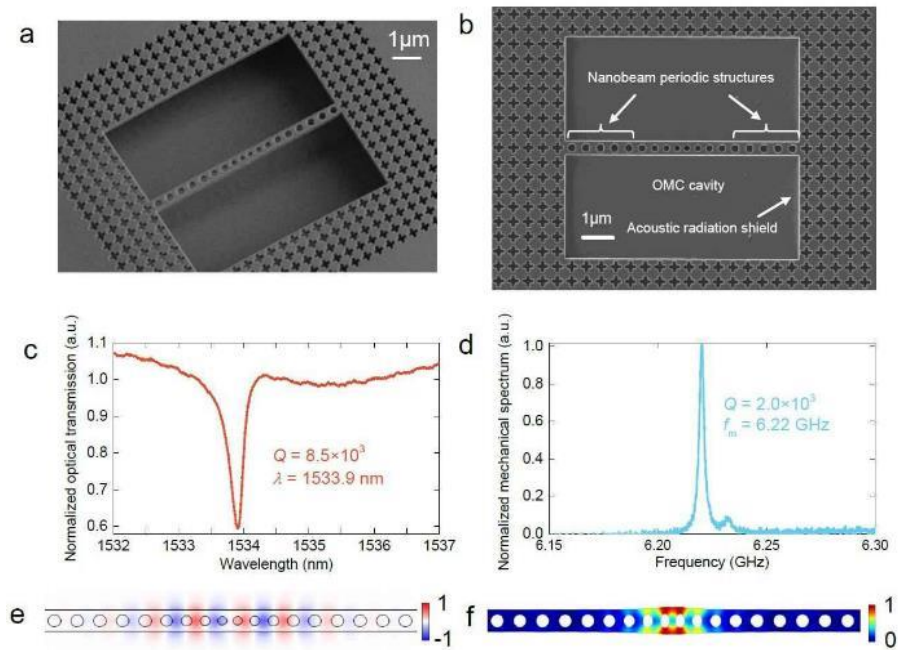

**Figure 3.** (**a**) Oblique view and (**b**) top view of the scanning electron microscope (SEM) image of the OMC cavity with an acoustic radiation shield. (**c**) Optical and (**d**) mechanical spectra of the OMC cavity. (**e**)

Transverse electric component of the cavity optical mode. (**f**) Displacement of the cavity mechanical mode.

**Simultaneous photon and phonon lasing**

We demonstrate simultaneous photon and phonon lasing using the fabricated OMC cavity in silicon in the normal regime with $\kappa \gg \Gamma$. As mentioned above, for an optomechanical cavity with $\kappa \gg \Gamma$, low-noise mechanical motion can be excited by a high-noise pump light. Thus, this regime has considerable potential for low-phase-noise applications. When we blue-detune the pump light with a linewidth of $2.0\times10^2$ kHz and increase its power, pronounced thresholds for both photons and phonons can be observed. Fig. 4a shows the normalised number of phonons in the cavity as a function of the coupled optical power. Here, the number of phonons is normalised to the number of phonons excited by the thermal environment, which is $1.01\times10^3$ at room temperature. Owing to the effective confinement of the optical and mechanical modes and the strong interaction between them in the OMC cavity, the threshold of the reduced optical power is 503 µW. Beyond this threshold, the number of phonons in the mechanical mode increases sharply, as shown in Fig. 4a, and the mechanical linewidth can be suppressed from 3.2 MHz to 5.2 kHz, as shown in Fig. 4b, corresponding to an effective mechanical Q-factor of up to $1.2\times10^6$. We further analyse the suppressed mechanical linewidth and find that it follows the ST limit form as shown in Fig. 4b (the details of the calculation of the ST-limit line are given in Supplementary Information section S10). The spectra of the spontaneous and stimulated phonon and scattering photon are shown in Fig. 4b. These experimental results are in good agreement with the aforementioned conclusion that the mechanical-lasing linewidth is much narrower than the linewidth of the pump laser.

Because the mechanical motion is characterised by the optical beat note between the scattering light and the pump light, the power of the scattering light can also be inferred (see Supplementary Information section S8). Fig. 4a shows the power of the scattering light versus the coupled optical power. Similar to the phonon number, the power of the scattering light increases sharply when the reduced pump power is beyond the threshold. In addition, because the linewidth of the beat signal changes from 3.2 MHz to 5.2 kHz and the linewidth of the pump laser itself is $2.0\times10^2$ kHz, it can be inferred that the linewidth of the scattering light changes from around 3.2 MHz to $2.0\times10^2$ kHz, which is consistent with our theoretical prediction outlined above. From the pronounced thresholds, increased power, and narrowed linewidth, we conclude that simultaneous photon and phonon lasing occurs beyond the threshold in our silicon OMC cavity. This experimental result can also be predicted by our analytical model of simultaneous photon and phonon lasing, which shows an identical threshold value of $|a_0| \geq \sqrt{\kappa\Gamma}/(2g_0)$ (see Supplementary Information S1-2). We attribute the phenomenon of simultaneous photon and phonon lasing in the cavity optomechanical system to the cavity-enhanced interaction of photons and phonons.

In addition, we measured the phonon linewidth excited by a pump laser with a linewidth of up to 50 MHz. The measured spectrum of the pump photon is shown in Fig. 3d. Although the linewidth of the pump is much greater than the intrinsic linewidth of the mechanical motion of 3.2 MHz, the linewidth of the excited phonon decreases to 5.4 kHz, which is four orders of magnitude narrower than the linewidth of the pump light. This is in good agreement with the simulation prediction shown in Fig. 2b, i.e. the ultra-low-noise mechanical signal can be excited even with a high-noise pump light, whereas the linewidths of the scattering photon and the pump light are similar.

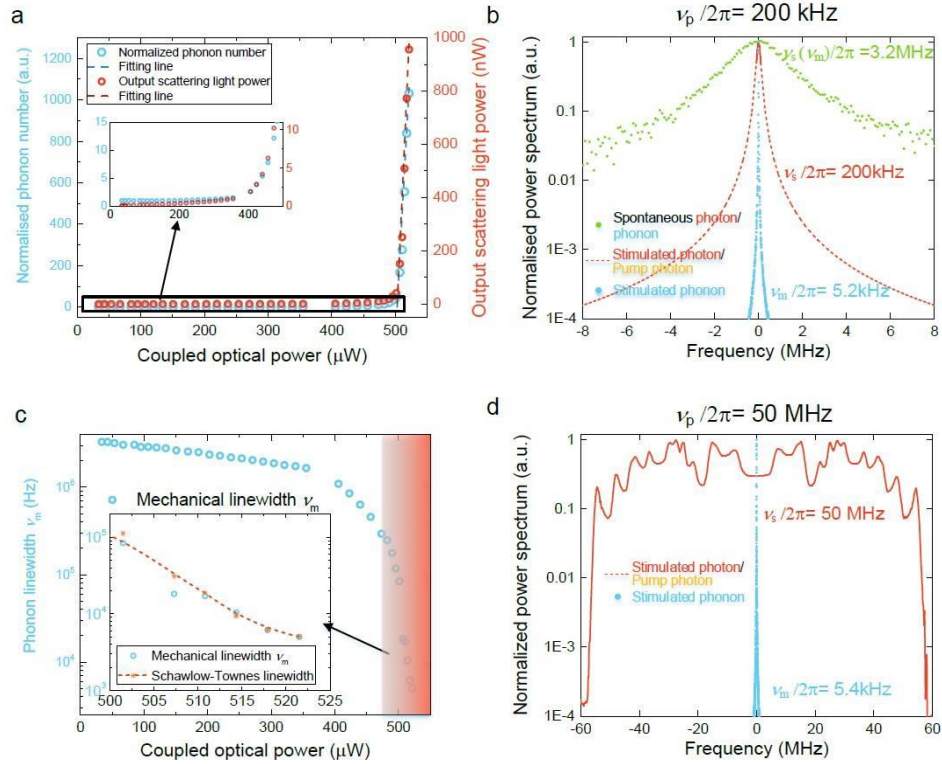

**Figure 4.** Experimental demonstration of simultaneous photon and phonon lasing in the silicon-based OMC cavity. (**a**) Normalised number of phonons (blue) and output scattering light power (red) as a function of the coupled optical power. The inset shows the data below the threshold. (**b**) Normalised power spectra of the spontaneous photon and phonon (green), stimulated photon (red), and stimulated phonon (blue) with a pump linewidth of 200 kHz. (**c**) Phonon linewidth as a function of the coupled optical power; the lasing zone is marked as a red gradient zone. The inset shows the calculated ST limit as a function of the coupled optical power. (**d**) Normalised power spectra of the stimulated photon (red) and the stimulated phonon (blue) with a pump linewidth of 50 MHz.

# Discussion and conclusions

Our theoretical analysis explained the dynamics underlying the lasing behaviour of an cavity optomechanical system. For systems in the normal regime with $\kappa \gg \Gamma$, a high-noise pump laser can excite an ultra-low-noise mechanical oscillation, and the low-noise signal can be directly detected by a photon detector from the high-noise pump laser without introducing additional noise. Moreover, the behaviour of the cavity optomechanical system is distinctly different from that of the Brillouin system in the reversed regime with $\kappa \ll \Gamma$, because the jitter in the equilibrium position is considerable and it influences the system significantly. The linewidth of the scattering photon is broader than the linewidth of the pump laser. However, the phonon linewidth is consistent with that of stimulated Brillouin scattering and it follows Eq. (9).

For experimental verification, we used a silicon-based OMC cavity with an acoustic radiation shield. We demonstrated both silicon-based photon and phonon lasers with phonon lasing at 6.22 GHz and scattering photon lasing at a wavelength of 1533.9 nm. The measured narrow linewidths for photon and phonon lasing were consistent with our theoretical predictions.

Our findings can pave the way towards photon and phonon lasers based on silicon, and they could enable researchers to meet the pressing need for new laser technologies for silicon-based photonic devices. Harnessing the coherence of photons and phonons in silicon devices could facilitate new applications in sensing, metrology, spectroscopy, and signal processing, thereby advancing the use of highly coherent sources.

Notably, our optomechanical-cavity-based scheme offers a footprint of only 120 μm$^2$, i.e. around three orders of magnitude smaller than that required by the Brillouin system[9], which requires a long optical path of several centimetres. Moreover, owing to the strong photon-phonon interaction in the OMC cavity, the threshold pump power required to achieve simultaneous lasing is only 503 μW, which is lower than that of the Brillouin laser (7.5 mW) by an order of magnitude[9]. Our OMC system is compatible with silicon integrated devices, and it may provide an elegant approach for complementing stimulated Brillouin scattering, which is not only the strongest and most tunable nonlinear interaction with considerable potential for high-coherence and ultra-low-noise oscillation but also particularly challenging to achieve on a silicon platform.

# Methods

**Simulation of the lasing dynamics**

The differential evolution equations of the optomechanical systems were solved numerically using the

Runge–Kutta method. The time step was set to 0.01 of the mechanical period. To emulate the Gaussian noise such as that due to the thermal force as well as the phase noise of the pump light, Gaussian-distributed stochastic quantities were introduced that were invariant in the same step and changed between different steps. To obtain the power spectra in the frequency domain, the optical amplitude and displacement in the time domain were Fourier-transformed, and the square norms were applied.

**Fabrication of the lasing cavity**

The patterns of the cavities were first defined by electron beam lithography and then transferred to a silicon-on-insulator (SOI) wafer with a device layer of 220 nm by inductively coupled plasma etching. Subsequently, the buried oxide layers beneath the cavities were removed by wet etching with buffered hydrofluoric acid to form the suspended structures.

**Measurements**

A tunable laser light source was used. The cavities were optically coupled using a tapered fibre, and the output was split into two channels. One port was connected to a low-frequency (kHz) optical power monitor, from which the optical spectrum was obtained by sweeping the wavelength of the laser. The other port was connected to a high-frequency (12.5 GHz) optical receiver, and its output was connected to an electrical spectrum analyser to obtain the mechanical spectra (see Supplementary Information section S5).

To conduct the lasing experiment, we blue-detuned the tunable laser, i.e. we set the wavelength of the laser below the cavity wavelength. To eliminate the effect of the optical frequency shift of the cavity owing to thermal noise, the wavelength of the tunable laser was tuned to deliver the highest power in the electrical spectrum analyser for each input power, ensuring a fixed detuning between the pump and the cavity (see Supplementary Information section S7). For example, because the ratio between the mechanical frequency and the total optical decay rate ($\Omega_m/\kappa$) was 0.21 for the OMC cavity, the largest signal from the electrical spectrum analyser was obtained when the ratio between the optical detuning and the total optical decay rate ($\Delta/\kappa$) was 0.31 (see Supplementary Information section S6).

**Code availability**

The custom code used in this study is available from the corresponding author upon reasonable request.

**Data availability**

The supporting data for the plots in this paper and other findings of this study are available from the corresponding author upon reasonable request.

**Acknowledgements**

The authors would also like to thank Dr. Di Qu and Mr. Guoren Bai of Innovation Center of Advanced Optoelectronic Chip and Institute for Electronics and Information Technology in Tianjin, Tsinghua University for their help with device fabrication. This work was supported by the National Key R&D Program of China under Contracts No. 2017 YFA0303700, the National Natural Science Foundation of China (Grant No. 61775115, 91750206, 61575102, and 61621064), the Opened Fund of the State Key Laboratory on Integrated Optoelectronics (No. IOSKL2016KF01), and Beijing Innovation Center for Future Chips, Tsinghua University.

**Author Contributions**

K.C., Z.H. and J.X. conceived the study. J.X. and K.C. performed the theoretical analysis, J.X. and Z.H. conducted the fabrication and measurement, and K.C. and Y.H. analyzed the results. K.C., Z.H. and J.X. wrote the paper. X.F., F.L., W.Z. discussed the results and reviewed the manuscript.

# Supplementary Information

## S1. Derivation of lasing threshold

### S1-1. Linearization of evolution equations

In cavity optomechanics, the total light amplitude can be considered as the summation of the steady-state term due to the optical cavity and the scattering term due to the optomechanical interaction, i.e.

$$a = a_0 + \delta a. \tag{S1}$$

Here, $a$ is the steady-state optical amplitude in the cavity and $\delta a$ represents the scattering term. Thus, the interaction term in the semi-classical regime, i.e. $-\hbar g_0 a^* a (b^* + b)$, will be

$$-\hbar g_0 \left( (a_0^* \delta a + a_0 \delta a^*)(b^* + b) + \delta a^* \delta a (b^* + b) \right) \tag{S2}$$

We consider that the origin of the mechanical oscillation is affected by the steady-state optical field. In Eq. (S2), $b$ stands for the normalized complex amplitude of the mechanical motion, and the displacement can be expressed as $x = x_{zpf}(b^* + b)$, where $x_{zpf}$ is the zero-point fluctuation amplitude of the mechanical oscillator[1].

When $\delta a$ is relatively small, a common method to deal with the optomechanical cavity system is to ignore the term $\delta a^* \delta a$ in Eq. (S2). Thus, Eqs. (1) and (2) in the main text become[1]

$$\delta \dot{a} = (i\Delta - \frac{\kappa}{2})\delta a + i \frac{g_0}{x_{zpf}} x a_0, \tag{S3}$$

$$m\ddot{x} = -m\Omega_m^2 x - m\Gamma \dot{x} + \hbar \frac{g_0}{x_{zpf}}(a_0^* \delta a + a_0 \delta a^*) + f_{th}. \tag{S4}$$

If the normalized complex amplitude of the mechanical motion ($b$) is taken as a variable, Eq. (S4) becomes

$$\dot{b} = \left(-i\Omega_m - \Gamma/2\right)b + ig_0(a_0^* \delta a + a_0 \delta a^*) + \sqrt{\Gamma} b_{th}, \tag{S5}$$

where $b_{th} = \sqrt{k_B T / \hbar \Omega_m} \dot{B}_C$ is the fluctuation due to the thermal environment and $\dot{B}_C$ is the standard complex Wiener process. Note that owing to the large amplitude of $\delta a$, these linearized equations are not suitable for studying the dynamic properties of simultaneous photon and phonon lasing, as explained in the main text.

## S1-2 Threshold condition of simultaneous photon and phonon lasing

Below the laser threshold, if the pump light is blue-detuned relative to the optical cavity, the scattering photon and phonon (mechanical motion) will be enhanced. According to the photon picture (studying the generation and annihilation of scattering photon), in a vibration period of a mechanical oscillator, if the scattering of the pump owing to a small mechanical motion compensates the dissipation of the scattering photon, the corresponding critical pump power can be considered as the threshold condition of (scattering) photon lasing. Meanwhile, according to the phonon picture (studying the generation and annihilation of scattering phonon), when the generated rate of the phonon exactly compensates the dissipation rate, the corresponding critical pump power can be considered as the threshold condition of phonon lasing.

Nonlinear effects can be ignored when the system is below the threshold. Hence, it is valid to use the linearized equations (S3) and (S5) to derive the lasing threshold condition. First, we assume that there is a slight vibration at the beginning of the mechanical mode, which can be set to $x = x_0 \sin(\Omega_m t)$. In the derivation of the threshold condition, we do not consider any random noise. Thus, $x_0$ is a real number and there is no random phase term in the sinusoidal function. If we keep only the low-order scattering terms $a_1$ and $a_{-1}$ (no additional random phase noise is included here), the steady-state solution of the scattering light can be obtained from the linearized evolution equations, and it is denoted as follows:

$$a_1 = \frac{g_0 a_0 x_0}{2 x_{zpf}} \frac{1}{-i(\Delta - \Omega_m) + \kappa/2}, \tag{S6}$$

$$a_{-1} = -\frac{g_0 a_0 x_0}{2 x_{zpf}} \frac{1}{-i(\Delta + \Omega_m) + \kappa/2}. \tag{S7}$$

**Photon picture:** In a sideband resolved regime, we can ignore $a_{-1}$ and put $\delta a = a_1 e^{i\Omega_m t}$ into (S4). This leads to a mechanical motion $x = \frac{2\hbar g_0}{m x_{zpf} \Omega_m \Gamma} \text{Im}\left(a_0^* a_1 e^{i\Omega_m t}\right)$, which is driven by the radiation pressure term $a_0^* \delta a$. We substitute $x$ into (S3). When the system is at the critical point of lasing, the dissipation of the scattering light should be compensated. Accordingly, the threshold condition of phonon lasing can be expressed as

$$|a_0| \geq \frac{\sqrt{\kappa \Gamma}}{2 g_0} \tag{S8}$$

**Phonon picture:** According to the energy conservation condition, the phonon generating rate is $\eta_{sc} = \kappa\left(|a_1|^2 - |a_{-1}|^2\right)$. Thus, the power obtained by the mechanical oscillator is $P = \eta_{sc} \hbar \Omega_m$. We substitute this expression into (S6) and (S7) and we use both the sideband resolved condition $\Omega_m \gg \kappa$ and the detuning condition $\Delta \approx \Omega_m$. Thus, the threshold expression can be derived as $|a_0| \geq \sqrt{\kappa \Gamma}/(2 g_0)$, which is completely consistent with the threshold condition (S8) obtained in the photon picture. Note that the threshold condition can be applied to both $\Gamma \gg \kappa$ and $\Gamma \ll \kappa$ regimes because a comparison of the linearized equations in these two regimes shows that there are indeed no differences in form. Therefore, under the sideband resolved condition, the scattering photon and phonon have the same lasing threshold of $\sqrt{\kappa \Gamma}/(2 g_0)$. According to previous studies[2,3], photons and phonons cannot lase simultaneously. Based on our derivation for both photon and phonon pictures, we show a very significant result that simultaneous phonon and photon lasing can occur in optomechanical cavities.

## S2. Dynamics of optomechanical systems with κ < Γ when ignoring $\delta a^* \delta a$

Based on the linearization of the evolution equations, the scattering photon ($\delta a$) and the mechanical displacement ($x$), expressed as $\delta a = a_1 e^{i\Omega_m t} + a_{-1} e^{-i\Omega_m t}$ and $x = x_{zpf}(b_1 e^{i\Omega_m t} + b_1^* e^{-i\Omega_m t})$, can be substituted into Eqs. (S3)–(S5), and the evolution equations can be simplified as

$$\dot{a}_0 = -\frac{\kappa}{2}a_0 + i\Delta a_0 + ig_0(b_1^* a_{-1} + b_1 a_1) + \sqrt{\kappa_{ex}}\, s_{in}, \tag{S9}$$

$$\dot{a}_1 = [-\frac{\kappa}{2} + i(\Delta - \Omega_m)]a_1 + ig_0 b_1^* a_0, \tag{S10}$$

$$\dot{a}_{-1} = [-\frac{\kappa}{2} + i(\Delta + \Omega_m)]a_{-1} + igb_1 a_0, \tag{S11}$$

$$\dot{b}_1 = -\frac{\Gamma}{2}b_1 + ig_0(a_0^* a_{-1} + a_0 a_1^*) + \sqrt{\Gamma}\, b_{th}. \tag{S12}$$

Here, $a_1$, $a_{-1}$, and $b_1$ are the slowly varied amplitudes, and the variation of $a_0$ is considered as well.

We simulate the lasing dynamics according to Eqs. (S9)–(S12) for the optomechanical system with the same parameters as those in Fig. 2(c) in the main text (without considering the thermal noise), i.e. optical decay rate $\kappa/2\pi$ = 6.5 MHz, mechanical decay rate $\Gamma/2\pi$ = 197 MHz, and pump linewidth $\nu_p/2\pi = 100$ kHz. The linewidths of the scattering photon and phonon are shown in Fig. S1(a), and their phase fluctuations are shown in Fig. S1(b). As shown in Fig. S1(b), the phase of the phonon follows the phase fluctuation of the pump and the phase of the scattering photon is much more stable. Consequently, the linewidth of the scattering photon is much narrower than that of the pump light, which is around 1 kHz.

These results are obtained with the nonlinear term ignored and the conclusion on the linewidth for the scattering photon is different from that with the dynamics where the nonlinear term is preserved. For the latter, the linewidth of the scattering photon is broader than that of the pump, as shown in Fig. 2c in the main text. This implies that the nonlinear term $\delta a^* \delta a$ of the optomechanical cavity system causes the linewidth of the scattering photon to become narrower than that of the pump. Therefore, linearization approximation in the existing study is invalid and it could lead to incorrect conclusions[3].

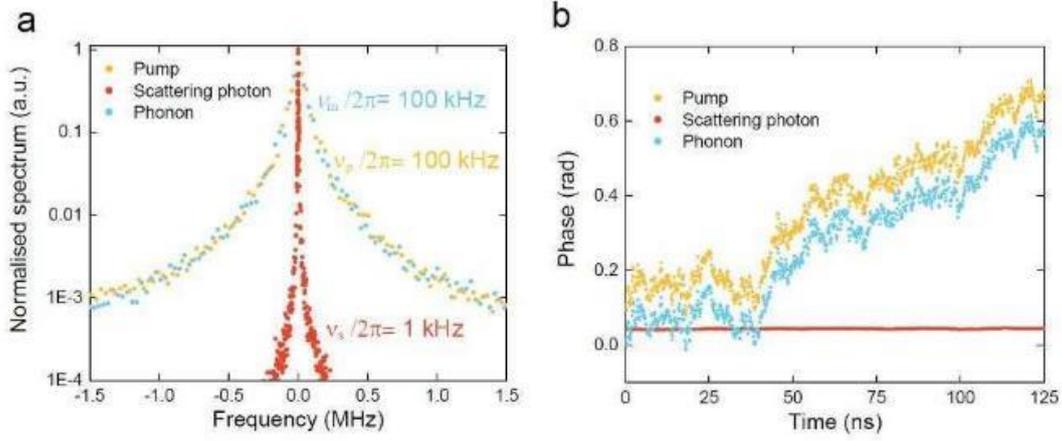

**Figure S1.** Simulation results of an optomechanical system with κ < Γ when $\delta a^*\delta a$ is ignored. (**a**) Spectra of the pump, scattering photon, and phonon after lasing. (**b**) Phase fluctuation of the pump, scattering photon, and phonon after lasing for the system represented in **a**.

## S3. Limit cycle theory and linewidth analysis

### S3-1 Evolution equations using a general limit cycle theory without linearization approximation

When the optomechanical system is a self-sustained oscillating (phonon lasing) system, it should be in a stable limit cycle state[4], i.e. each parameter can be expanded as a Fourier series. Here, only the first harmonic term of the mechanical mode needs to be retained, since multiply frequency of the fundamental mechanical mode is not resonant modes thus suffers from considerable dissipation. According to this scenario, the light field ($a$) and the mechanical motion ($b$) can be expressed as

$$a = \sum_n \alpha_n e^{in\cdot\Omega_m t}, \tag{S13}$$

$$b = B_c + Be^{-i\Omega_m t}. \tag{S14}$$

Here, $\alpha_n$ is the normalized complex amplitude of photons with different frequencies $\omega_n$ in the cavity (i.e. $\alpha_n = A_n e^{i\varepsilon_n}$), $B_c$ represents the equilibrium position of the mechanical vibration, and $B$ is the normalized complex amplitude of the mechanical vibration (including the amplitude $B_0$ and the phase $\phi$, i.e. $b = B_c + B_0 e^{i\phi} e^{-i\Omega_m t}$). We substitute Eqs. (S13) and (S14) into Eqs. (1) and (2) in the main text, and we get the following series expansion formula:

$$\sum_n (\dot{\alpha}_n + in\Omega_m \alpha_n) e^{in\cdot\Omega_m t} = (i\Delta - \kappa/2)\sum_n \alpha_n e^{in\cdot\Omega_m t} + i2g_0 \text{Re}[B_c]\sum_n \alpha_n e^{in\cdot\Omega_m t}$$
$$+ ig_0(Be^{-i\Omega_m t} + B^* e^{i\Omega_m t})\sum_n \alpha_n e^{in\cdot\Omega_m t} + \sqrt{\kappa_{ex}} s_{in}, \tag{S15}$$

$$\dot{B}_c + (\dot{B} - i\Omega_m B)e^{-i\Omega_m t} = -(i\Omega_m + \Gamma/2)(B_c + Be^{-i\Omega_m t})$$
$$+ ig_0 \sum_n \alpha_n e^{in\cdot\Omega_m t} \times \sum_k \alpha_k^* e^{-ik\cdot\Omega_m t} + \sqrt{\Gamma} b_{th}. \tag{S16}$$

In the above-mentioned equations, the evolutions of modes of different frequencies are considered to be independent of each other. Thus, the above-mentioned equations can be further decomposed into evolution equations for $\alpha_n$, $B$, and $B_c$.

$$\dot{\alpha}_n = \left[i\left(\Delta + 2g_0 \operatorname{Re}[B_c] - n\Omega_m\right) - \frac{\kappa}{2}\right]\alpha_n + ig_0\left(B\alpha_{n+1} + B^*\alpha_{n-1}\right) + \sqrt{\kappa_{ex}}|s_{in}|e^{i\varepsilon_p}\delta_{n,0}, \tag{S17}$$

$$\dot{B} = -\frac{\Gamma}{2}B + ig_0\sum_n \alpha_n \alpha_{n+1}^* + \sqrt{\Gamma}b_{th}e^{i\Omega_m t}, \tag{S18}$$

$$\dot{B}_c = -(i\Omega_m + \frac{\Gamma}{2})B_c + ig_0\sum_n |\alpha_n|^2. \tag{S19}$$

Here, $\Delta + 2g_0\operatorname{Re}[B_c]$ is the equivalent detuning $\tilde{\Delta}$, indicating the change in the optical resonant frequency caused by the change in the mechanical equilibrium position. To analyse the dynamics of the amplitude and phase, respectively, we decompose $\alpha_n$ and $B$ into the form of the amplitude component and phase component, i.e. $\alpha_n = A_n e^{i\varepsilon_n}$ and $B = B_0 e^{i\phi}$. Using the evolution equations (S17) and (S18), we can further derive the evolution equations of $A_n$, $\varepsilon_n$, $B_0$, and $\phi$:

$$\dot{A}_n = -\frac{\kappa}{2}A_n - g_0 B_0 \left[A_{n+1}\sin(\phi+\varepsilon_{n+1}-\varepsilon_n) - A_{n-1}\sin(\phi+\varepsilon_n-\varepsilon_{n-1})\right] + \sqrt{\kappa_{ex}}|s_{in}|\cos(\varepsilon_p - \varepsilon_n)\delta_{n,0}, \tag{S20}$$

$$\dot{B}_0 = -\frac{\Gamma}{2}B_0 + g_0\sum_n A_n A_{n+1}\sin(\phi+\varepsilon_{n+1}-\varepsilon_n) + \sqrt{\Gamma}b_{th}\cos(\Omega_m t - \phi), \tag{S21}$$

$$\dot{\varepsilon}_n = \tilde{\Delta} - p\Omega_m + g_0 B_0 \frac{A_{n+1}}{A_n}\cos(\phi+\varepsilon_{n+1}-\varepsilon_n) + g_0 B_0 \frac{A_{n-1}}{A_n}\cos(\phi+\varepsilon_n-\varepsilon_{n-1}) + \frac{\sqrt{\kappa_{ex}}|s_{in}|}{A_n}\sin(\varepsilon_p - \varepsilon_n)\delta_{n,0}, \tag{S22}$$

$$\dot{\phi} = \frac{g_0}{B_0}\sum_n A_n A_{n+1}\cos(\phi+\varepsilon_{n+1}-\varepsilon_n) + \frac{\sqrt{\Gamma}}{B_0}b_{th}\sin(\Omega_m t - \phi), \tag{S23}$$

$$\dot{B}_c = -(i\Omega_m + \frac{\Gamma}{2})B_c + ig_0\sum_n A_n^2. \tag{S24}$$

In general, these five equations are coupled to each other and thus need to be solved together. For lasing dynamics, as the amplitude noise is suppressed[5], we can use $\dot{A}_n \ll A_n$ and $\dot{B}_0 \ll B_0$ to reduce these equations. Accordingly, we can assume that the resonant amplitude is approximately a constant; thus, we can replace the original random variable with its expectation in these equations, i.e. $A_n \approx \langle A_n \rangle$ and $B_0 \approx \langle B_0 \rangle$. Then, the evolution equations of phase $\varepsilon_n$ and $\phi$ can be decoupled from the evolution equations of amplitude. Under this condition, we derive the 'phase reduction equations'[6] for an optomechanical system in the phonon lasing regime:

$$\dot{\varepsilon}_n = \tilde{\Delta} - n\Omega_m + g_0\langle B_0 A_{n+1}/A_n\rangle\cos(\phi+\varepsilon_{n+1}-\varepsilon_n)$$

$$+ g_0\langle B_0 A_{n-1}/A_n\rangle\cos(\phi+\varepsilon_n-\varepsilon_{n-1}) + \frac{\sqrt{\kappa_{ex}}|s_{in}|}{\langle A_p\rangle}\sin(\varepsilon_p - \varepsilon_n)\delta_{n,0}, \quad p \in Z \tag{S25}$$

$$\dot{\phi} = g_0\sum_n\langle A_n A_{n+1}/B_0\rangle\cos(\phi+\varepsilon_{n+1}-\varepsilon_n) + \frac{\sqrt{\Gamma}}{\langle B_0\rangle}b_{th}\sin(\Omega_m t - \phi). \tag{S26}$$

The lasing linewidth originates from the pump phase noise[7] such that the angular frequency linewidth of the phonon is equal to the variance of $\dot{\phi}$ and the angular frequency linewidth of the p-order scattering light is equal to the variance of $\dot{\varepsilon}_n$, which means that the linewidth is determined by the randomness on the right-hand side of these equations. In summary, there are four random items in Eqs. (S24) and (S25): $\tilde{\Delta}$, $\phi$, $\varepsilon_n$, and $\varepsilon_p$. Here, $\tilde{\Delta}$ represents the equivalent cavity detuning, which is usually considered to be a constant[1]. However, we do not make this assumption and we first analyse the randomness of $\tilde{\Delta}$. The detailed analysis of the other random variables $\phi$, $\varepsilon_n$, and $\varepsilon_p$ will be discussed in section S3-3.

To analyse the random variable $\tilde{\Delta}$, we separate the variation (i.e. $\delta B_c = B_c - \langle B_c \rangle$) from Eq. (S24) and get $\delta B_c$ as

$$\delta \dot{B}_c = -(i\Omega_m + \frac{\Gamma}{2})\delta B_c + i2g_0 \sum_n A_n \delta A_n . \tag{S27}$$

Based on the assumptions that $\delta \dot{B}_c \ll \Omega_m \delta B_c$ and $\Gamma \ll \Omega_m$, $\delta B_c$ can be expressed as

$$\delta B_c = (2g_0 / \Omega_m) \sum_n A_n \delta A_n . \tag{S28}$$

Thus, the random fluctuation of the equivalent frequency detuning can be obtained as

$$\delta \tilde{\Delta} = 2g_0 \operatorname{Re}[\delta B_c] = (4g_0^2 / \Omega_m) \sum_n A_n \delta A_n . \tag{S29}$$

It can be seen that the magnitude of the fluctuation of equivalent detuning $\tilde{\Delta}$ is related to the field amplitude in the cavity. As the energy conservation relationship between the scattering photon and the phonon is maintained, if the number of phonons is assumed to be a known variable, the ratio of the dissipation rate can be used to characterize the number of scattering photons (considering only three main optical modes, i.e. Stokes light $A_1$, anti-Stokes light $A_{-1}$, and pump light $A_0$):

$$B_0^2 / (A_1^2 - A_{-1}^2) = \kappa / \Gamma \tag{S30}$$

When the anti-Stokes light $A_1$ is ignored in the case of blue-detuning, we can use the reduced expression $B_0 / A_1 = \sqrt{\kappa / \Gamma}$ to simplify Eq. (S29) as

$$\delta \tilde{\Delta} = 2g_0 \operatorname{Re}[\delta B_c] = (4g_0^2 / \Omega_m)(B_0 \delta B_0 + A_0 \delta A_0) \tag{S31}$$

According to the analysis on page 6 of the main text, it is easily found that in the reversed regime, the fluctuation of $A_1$ will lead to a large increase in the variance $\operatorname{var}(\delta \tilde{\Delta})$. Thus, the linewidth of the scattering photon will be strongly affected by this term. However, in the normal regime, the corresponding fluctuation is extremely weak; hence, in this regime, it is reasonable to regard $\tilde{\Delta}$ as a constant. To illustrate the difference in the fluctuation of $A_1$ between the normal regime and the reversed regime, we plot the numerical simulation results of these two distinct regimes in Fig. S2a and Fig. S2b. From Eq. (7) in the main text, the pump phase noise causes fluctuation in the light intensity ($A_1$), which leads to jitter in the equilibrium position $\delta B_c$ (or, equivalently, jitter in the effective detuning $\delta \tilde{\Delta}$). This effect becomes considerable in the reversed regime because the jitter in the light intensity owing to the pump phase noise is remarkable.

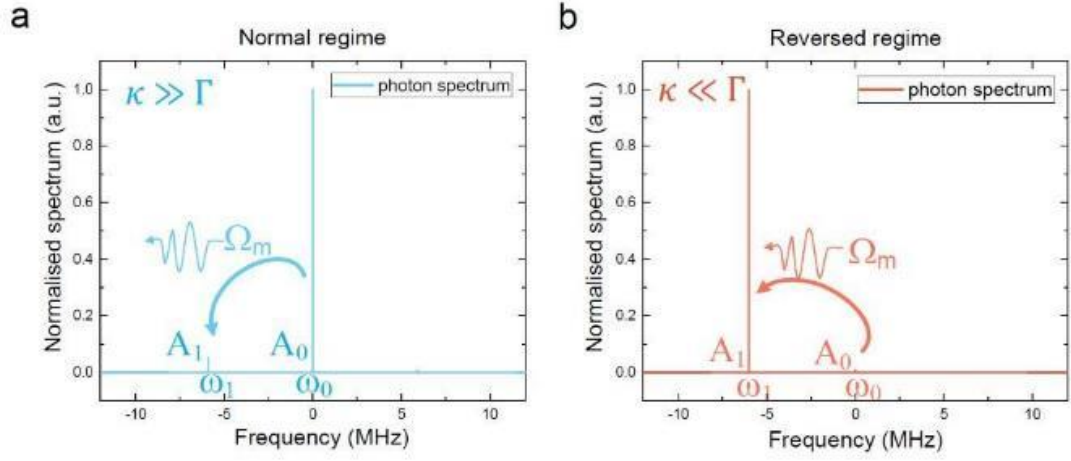

**Figure S2. (a)** Simulation results of the entire optical spectrum in the normal regime. $A_1$ and $A_0$ represent the electric amplitude of the scattering light and pump light, respectively. It is obvious that the pump light is much stronger than the scattering light. **(b)** Corresponding simulation results in the reversed regime. Compared to the pump, there is a dominant spectrum line, i.e. the scattering light.

## S3-2 Phase synchronization (matching) condition

When the system is in a self-sustained oscillating state, the phase diagram for the periodic variables should be restricted to a well-defined limit cycle[4]. When we ignore the all the potential noise in the system, the amplitude of the harmonic oscillator should be a constant. As $\Gamma B_0 / 2 = g_0 \sum_n A_n A_{n+1} \sin(\phi + \theta_n)$ (derived from Eq. (S21), $\theta_n = \varepsilon_{n+1} - \varepsilon_n$ is the phase of the beat note) and $\Gamma B_0 / 2$, $A_n$ are constants, $\phi + \theta_n$ must be a constant phase $\varphi_n$, which means that the phases $\phi$ and $\theta_n$ are strictly synchronized. When the system has relatively weak noise, the trajectory of the phase will diffuse and form a phase diagram with probability distributions, which can be described by the Fokker–Plank equation[8]. We use this conclusion directly (i.e. considering all the periodic variables are still well restricted near their limit cycle trajectories), and by applying the slowly varying amplitude approximation with a small jitter $\delta B_0$ near its amplitude expectation $\langle B_0 \rangle$, we can derive the following relations from Eq. (S21):

$$\frac{\Gamma}{2} \langle B_0 \rangle = g_0 \sum_n \langle A_n A_{n+1} \rangle \sin(\varphi_n), \tag{S32}$$

$$\frac{\Gamma}{2} \delta B_0 = g_0 \sum_n \langle A_n A_{n+1} \rangle \cos(\varphi_n) \delta\varphi_n. \tag{S33}$$

Here, we ignore the contribution of $\delta A_n = A_n - \langle A_n \rangle$ to assume that the phase noise dominates. From Eq. (S33), we find that the phase disturbance ($\delta\varphi_n$) leads to random jitter in the amplitude ($\delta B_0$). Hence, when considering noise, there is no strict phase-synchronizing condition between the beat note phase $\theta_n$ and the mechanical motion (strict phase synchronizing means that $\phi+\theta_n$ must be a constant phase $\varphi_n$). Therefore, in the presence of noise, the strict phase synchronizing conditions will be relaxed to $\phi+\theta_n = \varphi_n + \delta\varphi_n$ ($\delta\varphi_n$ represents a small phase jitter).

### S3-3 Lasing linewidth in $\Gamma \ll \kappa$ and $\Gamma \gg \kappa$ regimes

Here, the analytical expressions for the lasing linewidth are derived under the sideband resolved condition such that only two optical modes and one acoustic mode need to be considered, which means that Eqs. (S25) and (S26) can be applied and simplified. By applying $B_0/A_1 = \sqrt{\kappa/\Gamma}$ (derived from Eq. (S30)) and the laser threshold condition expression (S8), we can derive the following inequalities (equations) in the sideband resolved regime:

$$g_0 A_0 \frac{B_0}{A_1} \geq \frac{\kappa}{2} \quad \text{or} \quad g_0 A_0 \frac{B_0}{A_1} = \frac{M\kappa}{2}, \tag{S34}$$

$$g_0 A_0 \frac{A_1}{B_0} \geq \frac{\Gamma}{2} \quad \text{or} \quad g_0 A_0 \frac{A_1}{B_0} = \frac{M\Gamma}{2}. \tag{S35}$$

Here, we define $M$ as a nonlinear factor, which represents the gain saturation of the optomechanical system after lasing.

By applying Eqs. (S34) and (S35) to the phase reduction equations (S25) and (S26), we can obtain the simplified phase evolution equations:

$$\dot{\phi} = \frac{M\Gamma}{2}\cos(\phi+\varepsilon_1-\varepsilon_0) + \frac{\sqrt{\Gamma}}{\langle B_0 \rangle} b_{th} \sin(\Omega_m t - \phi), \tag{S36}$$

$$\dot{\varepsilon}_1 = \tilde{\Delta} - \Omega_m + \frac{M\kappa}{2}\cos(\phi+\varepsilon_1-\varepsilon_0), \tag{S37}$$

$$\dot{\varepsilon}_0 = \tilde{\Delta} + g_0 \left\langle \frac{B_0 A_1}{A_0} \right\rangle \cos(\psi_0) + \frac{S_0}{\langle A_0 \rangle}\sin(\varepsilon_p - \varepsilon_0). \tag{S38}$$

The effective detuning $\delta\tilde{\Delta}$ is directly associated with the dissipation rate $\Gamma/\kappa$ (analysed in section S3-1). Next, we will discuss the lasing linewidth of the optomechanical system in the normal regime ($\Gamma \ll \kappa$) and the reversed regime ($\Gamma \gg \kappa$).

### S3-3-1 Normal regime ($\Gamma \ll \kappa$):

In the normal regime, as mentioned above, as the cavity decay rate is dominant, the intensity of the light field in the cavity is limited; hence, the fluctuation in the effective detuning $\tilde{\Delta}$ owing to the fluctuation in the equilibrium position can be ignored. We start from Eqs. (S36)-(S38) and we ignore the change in the equivalent detuning by replacing $\tilde{\Delta}$ with $\langle \tilde{\Delta} \rangle$. Then, the phase equations are

$$\dot{\phi} = \frac{M\Gamma}{2}\cos(\phi + \varepsilon_1 - \varepsilon_0) + \frac{\sqrt{\Gamma}}{\langle B_0 \rangle} b_{th} \sin(\Omega_m t - \phi), \tag{S39}$$

$$\dot{\varepsilon}_1 = \langle \tilde{\Delta} \rangle - \Omega_m + \frac{M\kappa}{2}\cos(\phi + \varepsilon_1 - \varepsilon_0), \tag{S40}$$

$$\dot{\varepsilon}_0 = \langle \tilde{\Delta} \rangle + g_0 \left\langle \frac{B_0 A_1}{A_0} \right\rangle \cos(\phi + \varepsilon_1 - \varepsilon_0) + \frac{\sqrt{\kappa_{ex}} |s_{in}|}{\langle A_0 \rangle} \sin(\varepsilon_p - \varepsilon_0). \tag{S41}$$

There are two types of noise sources in the system. One is the phase noise of the pump light (i.e. $\varepsilon_p$), and the other is the thermal noise of the mechanical oscillator (i.e. $b_{th}$). As the two types of noise come from different physical processes, there is no correlation between them. Thus, the phonon linewidth can be split into the Schawlow–Townes linewidth caused by thermal noise and the linewidth caused by the pump light. Owing to the phase synchronization condition, the phase between the pump photon, scattering photon, and phonon needs to satisfy two conditions:

A. The phonon phase ($\phi$) and the phase of the beat note ($\varepsilon_1 - \varepsilon_0$) must be synchronized, which is equivalent to $\phi + \varepsilon_1 - \varepsilon_0 = const\_phase + phase\_jitter$.

B. The injected light ($\varepsilon_p$) and the intra-cavity light field ($\varepsilon_0$) must be synchronized, which is equivalent to $\varepsilon_p - \varepsilon_0 = const\_phase + phase\_jitter$. Here, we define $\delta c_{inj}$ as the phase jitter and $\langle c_{inj} \rangle$ as the constant phase. Thus, the synchronization relation can be expressed as $\varepsilon_p - \varepsilon_0 = c_{inj}$.

As $\varepsilon_0$ includes nearly no contribution of thermal noise (determined entirely by the phase of the injected light) while $\phi$ includes the contribution of thermal noise, in accordance with the phase-synchronizing scenario ($\phi + \varepsilon_1 - \varepsilon_0 = const\_phase + phase\_jitter$), $\varepsilon_1$ must include a thermal noise component complementary to $\phi$, thereby maintaining the stability of the limit cycle. Accordingly, we decompose the phase $\varepsilon_0$ of the Stokes light into the contribution of the thermal noise $\varepsilon_{1th}$ and the contribution of the pump light $\varepsilon_{1p}$. At the same time, the phase $\phi$ of the phonon is also decomposed into the contribution of the thermal noise $\phi_{th}$ and the pump $\phi_p$. Following the above analysis, we have

$$\varepsilon_0 = \varepsilon_p - c_{inj}, \text{ (phase synchronization between injected light and pump light)} \tag{S42}$$

$$\phi = \phi_{th} + \phi_p, \text{ (noise of phonons comes from thermal noise and pump phase noise)} \tag{S43}$$

$$\varepsilon_1 = \varepsilon_{1th} + \varepsilon_{1p}, \text{ (noise of photon comes from thermal noise and pump phase noise)} \tag{S44}$$

$$\varepsilon_{1th} + \phi_{th} = \delta_{th}, \text{ (thermal noise phase of phonon and scattering photon should synchronize)} \tag{S45}$$

$$\phi_p + \varepsilon_{1p} - \varepsilon_0 = \langle \psi_0 \rangle + \delta_p. \text{ (The phase of the pump light, scattering light, and phonon should synchronize)}$$

$$\tag{S46}$$

Here, as the pump optical phase noise and thermal noise will lead to diffusion of the limit cycle, we define the phase jitter $\delta_{th}$, which comes from thermal noise, and the phase jitter $\delta_p$, which comes from the pump phase noise. Using (S42)–(S46), we get $\phi + \varepsilon_1 - \varepsilon_0 = (\phi_{th} + \varepsilon_{1th}) + (\phi_p + \varepsilon_{1p} - \varepsilon_0) = \delta_{th} + (\langle\psi_0\rangle + \delta_p)$. This relation can be used to simplify Eqs. (S39)–(S41):

$$\dot{\phi}_p + \dot{\phi}_{th} = \frac{M\Gamma}{2}\cos\left[\delta_{th} + (\langle\psi_0\rangle + \delta_p)\right] + \frac{\sqrt{\Gamma}}{\langle B_0\rangle}b_{th}\sin(\Omega_m t - \phi), \quad (S47)$$

$$\dot{\varepsilon}_{1p} + \dot{\varepsilon}_{1th} = \langle\tilde{\Delta}\rangle - \Omega_m + \frac{M\kappa}{2}\cos\left[\delta_{th} + (\langle\psi_0\rangle + \delta_p)\right], \quad (S48)$$

$$\dot{\varepsilon}_0 = \langle\tilde{\Delta}\rangle + g_0\left\langle\frac{B_0 A_1}{A_0}\right\rangle\cos\left[\delta_{th} + (\langle\psi_0\rangle + \delta_p)\right] + \frac{\sqrt{\kappa_{ex}}|s_{in}|}{\langle A_0\rangle}\sin(c_{inj}). \quad (S49)$$

Using the above-mentioned conclusions, we can decompose Eqs. (S39) and (S40) into two parts, which represent the dynamics of thermal noise ($b_{th}$) and the dynamics of pump phase noise ($\varepsilon_p$).

**1. Dynamics of thermal noise**

From Eq. (S47), we can derive the thermal noise dynamics. The corresponding stochastic differential equation can be expressed as

$$\dot{\phi}_{th} = \frac{\sqrt{\Gamma}}{\langle B_0\rangle}b_{th}\sin(\Omega_m t - \phi), \quad (S50)$$

where the impact of $\delta_{th}$ on $\dot{\phi}_{th}$ is ignored. According to the method used by Melvin Lax[9], when the phase diffusion is not severe, we can simply drop the random phase term $\phi$ on the right-hand side, i.e. $\dot{\phi}_{th} = \sqrt{\Gamma}b_{th}\sin(\Omega_m t)/B_0$. We can calculate the variance of $\dot{\phi}_{th}$ by integrating over time $0 \sim t$ and obtain

$$\left\langle[\phi(t) - \phi(0)]^2\right\rangle = \frac{\Gamma}{|B_0|^2}\int_0^t d\tau_2 \int_0^t d\tau_1 \sin(\Omega_m \tau_1)\sin(\Omega_m \tau_2)\langle b_{th}(\tau_1)b_{th}(\tau_2)\rangle$$

$$\approx \frac{1}{2}\frac{\Gamma}{|B_0|^2}n_{th}[t - \sin(2\Omega_m t)/(2\Omega_m)] \xrightarrow{t\to\infty} \frac{1}{2}\frac{\Gamma n_{th}}{|B_0|^2}t. \quad (S51)$$

From the above-mentioned formula, we can get the well-known Einstein diffusion constant $D = \Gamma n_{th}/(4|B_0|^2)$ and use the relationship $n = |B_0|^2$ to directly obtain the phonon linewidth due to the thermal noise:

$$v_{mT} = 2D = \frac{\Gamma n_{th}}{2n}. \quad (S52)$$

This formula is consistent with the ST linewidth formula for phonons in the optomechanical cavity system derived by Kerr J. Vahala[10].

Further, by using the phase matching condition of Eq. (S45), we find that the thermal noise of the phonon can be transmitted to the scattering photon, i.e.

$$\text{var}(\dot{\varepsilon}_{1\text{th}}) \approx \text{var}(\dot{\phi}_{\text{th}}) = v_{\text{mT}} = \frac{\Gamma n_{\text{th}}}{2n}. \tag{S53}$$

In summary, owing to the introduction of thermal noise, the phase of the scattering photon and the phase of the phonon are diffused, resulting in linewidth broadening. For thermal noise, the linewidth of the scattering photon and the broadening of the phonon linewidth are nearly uniform, and the linewidth broadening caused by such noise has a form that is similar to that of the ST linewidth formula for the laser oscillator[10]. When the number of phonons in the cavity is large, ultra-narrow linewidth can be obtained.

## 2. Dynamics of phase noise

From Eqs. (S47) and (S48), we can derive the phase noise dynamics. The corresponding stochastic differential equations can be expressed as

$$\dot{\phi}_{\text{p}} = \frac{M\Gamma}{2}\cos(\phi_{\text{p}} + \varepsilon_{1\text{p}} - \varepsilon_0), \tag{S54}$$

$$\dot{\varepsilon}_{1\text{p}} = \langle \tilde{\Delta} \rangle - \Omega_{\text{m}} + \frac{M\kappa}{2}\cos(\phi_{\text{p}} + \varepsilon_{1\text{p}} - \varepsilon_0). \tag{S55}$$

Here, the subscript p represents the phase. We also ignore the influence of $\delta_{\text{th}}$. According to the two above-mentioned equations, we can directly calculate the variance on the right-hand side of the equations. Here, we need to consider only the random terms, and other constant terms can affect only the expectation values. The derived variance can be expressed as

$$v_{\text{m\_p}} = \text{var}(\dot{\phi}_{\text{p}}) = \Gamma^2 \left(\frac{M}{2}\right)^2 \text{var}\left[\cos(\phi_{\text{p}} + \varepsilon_{1\text{p}} - \varepsilon_0)\right], \tag{S56}$$

$$v_{\text{s\_p}} = \text{var}(\dot{\varepsilon}_{1\text{p}}) = \kappa^2 \left(\frac{M}{2}\right)^2 \text{var}\left[\cos(\phi_{\text{p}} + \varepsilon_{1\text{p}} - \varepsilon_0)\right]. \tag{S57}$$

From Eqs. (S56)–(S57), we can get $\text{var}(\dot{\phi}_{\text{p}})/\text{var}(\dot{\varepsilon}_{1\text{p}}) = \Gamma^2/\kappa^2$. By using the phase synchronization condition of Eq. (S46), we can obtain the relationship between the linewidth of the pump, scattering photon, and phonon as follows:

$$\text{var}(\dot{\varepsilon}_0) \approx \text{var}(\dot{\varepsilon}_{1\text{p}} + \dot{\phi}_{\text{p}}) = (\kappa+\Gamma)^2 \left(\frac{M}{2}\right)^2 \text{var}\left[\cos(\phi_{\text{p}} + \varepsilon_{1\text{p}} - \varepsilon_0)\right]. \tag{S58}$$

From Eqs. (S56)–(S58), the influence of the phase noise of the pump on the scattering photon and phonon linewidth is derived:

$$v_{\text{s\_p}} = \frac{v_{\text{p}}}{(1+\Gamma/\kappa)^2}, \tag{S59}$$

$$v_{\text{m\_p}} = \frac{v_{\text{p}}}{(1+\kappa/\Gamma)^2}. \tag{S60}$$

Here, $v_{\text{p}}$ represents the angular linewidth of the pump. The two above-mentioned expressions give us the scattering photon linewidth and the phonon linewidth without considering thermal noise. As the thermal noise is not correlated with the phase noise of the pump light, the linewidth contribution can be directly added, and the final linewidth expression is provided in Eqs. (8) and (9) in the main text.

## S3-3-2 Reversed regime ($\Gamma \gg \kappa$):

According to the analysis of the randomness of the equivalent detuning $\tilde{\Delta}$ owing to the fluctuation of the equilibrium position in section S3-1, the random fluctuation of $\tilde{\Delta}$ in this regime is considerable, resulting in considerable linewidth broadening. To analyse the effect on the linewidths of the phonon and scattering photon, we proceed from Eqs. (S25) and (S26). Here, we ignore the thermal noise and add the linewidth originating from the thermal noise to the final linewidth formula (similar to the method used in the normal regime).

In the following derivation, we define $\theta_n = \varepsilon_{n+1} - \varepsilon_n$ as the phase of the beat note of the adjacent spectral line, $\varepsilon_c$ as the random phase caused by the jitter in the equivalent detuning $\tilde{\Delta}$ (i.e. $\dot{\varepsilon}_c = \delta\tilde{\Delta}$), and $\varepsilon'_n$ as the modified phase in which the effect of $\delta\tilde{\Delta}$ vanishes (i.e. $\varepsilon'_n = \varepsilon_n - \varepsilon_c$). Using Eqs. (S25) and (S26), we find that in the absence of thermal noise, the stochastic differential equations of random variables $\theta_n$ and $\phi$ can be written as

$$\dot{\phi} = g_0 \sum_n \langle A_n A_{n+1} / B_0 \rangle \cos(\phi + \theta_n), \tag{S61}$$

$$\dot{\theta}_n = -\Omega_m + g_0 \langle B_o / A_{n+1} \rangle [\langle A_n \rangle \cos(\phi + \theta_n) + \langle A_{n+2} \rangle \cos(\phi + \theta_{n+1})]$$

$$- g_0 \langle B_o / A_n \rangle [\langle A_{n+1} \rangle \cos(\phi + \theta_n) + \langle A_{n-1} \rangle \cos(\phi + \theta_{n-1})] + \frac{\sqrt{\kappa_{ex}} |s_{in}|}{\langle A_0 \rangle} \sin(\varepsilon_p - \varepsilon_0)(\delta_{n,-1} - \delta_{n,0}). \tag{S62}$$

In the stochastic differential equation (S25), for different parameters $n$, $\tilde{\Delta}$ is exactly the same, and $\theta_n$ is the phase of the beat note in the stochastic differential equation (S62); thus, the effect of $\tilde{\Delta}$ is eliminated. In addition, in Eq. (S62), as the thermal noise is ignored, the random phase of the phonon is related only to the phase $\theta_n$ of the beat note. Hence, we can conclude that the phonon linewidth (owing to the pump phase noise) is not affected by $\tilde{\Delta}$. According to this conclusion, when considering the linewidth characteristics of the reversed regime, although $\tilde{\Delta}$ is highly random, we can still assume that $\tilde{\Delta}$ is a constant when solving the phonon linewidth. In the sideband resolved regime, the resulting phase-reduction equation has exactly the same form as Eqs. (S39)–(S41). Thus, the final obtained phonon linewidth also has the same form, i.e.

$$\nu_m = \frac{\nu_p}{(1 + \kappa/\Gamma)^2}. \tag{S63}$$

According to Eq. (S63), in the reversed regime ($\Gamma \gg \kappa$), the linewidth of the phonon will be approximately equal to the linewidth of the pump. This conclusion is consistent with the previous theoretical analysis of stimulated Brillouin scattering[11]. As the phase noise of the optical field is directly related with $\tilde{\Delta}$, to analyse the linewidth of the scattering photon, we have to consider the randomness of $\tilde{\Delta}$. Using the definition $\varepsilon'_n = \varepsilon_n - \varepsilon_c$, we can separate the random jitter caused by $\tilde{\Delta}$ from Eqs. (S37)–(S38), and the following equations are obtained.

$$\dot{\phi} = \frac{M\Gamma}{2}\cos(\phi + \theta_0), \tag{S64}$$

$$\dot{\varepsilon}'_1 = \langle \tilde{\Delta} \rangle - \Omega_m + \frac{M\kappa}{2}\cos(\phi + \theta_0), \tag{S65}$$

$$\dot{\varepsilon}'_0 = \langle \tilde{\Delta} \rangle + g_0 \left\langle \frac{B_0 A_1}{A_0} \right\rangle \cos(\phi + \theta_0) + \frac{\sqrt{\kappa_{ex}}|s_{in}|}{\langle A_0 \rangle}\sin[(\varepsilon_p - \varepsilon_c) - \varepsilon'_0]. \tag{S66}$$

The above-mentioned equations can be regarded as the phase reduction equations in the case where a modified light field with a complex amplitude $S_0 e^{i(\varepsilon_p - \varepsilon_c)}$ is injected and serves as a new pump light. As the form of Eqs. (S64)–(S66) is the same as that of Eqs. (S39)–(S41), we can obtain the linewidths of the phonon and scattering photon by using the method applied in the normal regime ($\Gamma \ll \kappa$):

$$\text{var}(\dot{\varepsilon}'_1) = \frac{\text{var}(\dot{\varepsilon}_p - \dot{\varepsilon}_c)}{(1 + \Gamma/\kappa)^2}, \tag{S67}$$

$$\text{var}(\dot{\phi}) = \frac{\text{var}(\dot{\varepsilon}_p - \dot{\varepsilon}_c)}{(1 + \kappa/\Gamma)^2}. \tag{S68}$$

Comparing Eq. (S63) with Eq. (S68), as they both represent the phonon linewidth, we can obtain $v_p = \text{var}(\dot{\varepsilon}_p) = \text{var}(\dot{\varepsilon}_p - \dot{\varepsilon}_c)$. Thus, there must be a specific correlation between $\varepsilon_p$ and $\varepsilon_c$. After the analysis, we find that $\varepsilon_c$ should satisfy one of the following two conditions:

a) $\dot{\varepsilon}_c = 0$

Under this condition, the randomness of the equivalent detuning is considered to be weak and thus regarded as a constant; this case is invalid in the discussion of section (S3-1). However, if we follow this invalid assumption, we can get the (mistaken) linewidth of the scattering photon as

$$v_s = \frac{v_p}{(1 + \Gamma/\kappa)^2}. \tag{S69}$$

According to Eq. (S69), the phase noise of the pump has almost no effect on the linewidth of the scattering photon; thus, it can be narrowed considerably compared to the pump linewidth in the reversed regime. This is the so-called photon lasing regime[3]. However, in the practical optomechanical system, this conclusion is incorrect because the random fluctuations in the reversed regime are considerable. In the numerical simulation of section S2, by ignoring the high-order terms of the scattering light (equivalent to ignoring the randomness of $\tilde{\Delta}$), we can also obtain numerical simulation results that are consistent with Eq. (S69). However, for the high-order terms of the scattering photon, the simulation results suggest that the scattering photon linewidth cannot be narrowed relative to the pump (see Fig. 2 in the main text). Both the theoretical and numerical results confirm that the randomness of $\tilde{\Delta}$ in this regime cannot be ignored, and the linewidths of both the scattering photon and the phonon are broader than the linewidth of the pump light. The counterintuitive linewidth of the scattering photon originates from the jitter in the equilibrium position of the optomechanical cavity system and differs significantly from that of the Brillouin system.

b) $\dot{\varepsilon}_c = \dot{\varepsilon}_p + \dot{\varepsilon}_x$, $\mathrm{var}(\dot{\varepsilon}_x) = \mathrm{var}(\dot{\varepsilon}_p)$. Here, $\varepsilon_x$ is a random variable.

By applying this relation, $\mathrm{var}(\dot{\varepsilon}_p - \dot{\varepsilon}_c) = \mathrm{var}(\dot{\varepsilon}_p - \dot{\varepsilon}_p - \dot{\varepsilon}_x) = \mathrm{var}(-\dot{\varepsilon}_x) = v_p$ can be satisfied. Under this condition, the linewidth broadening of the scattering photon mainly comes from the fluctuation of $\tilde{\Delta}$, i.e. $v_s \approx \tilde{\Delta} \approx \mathrm{var}(\dot{\varepsilon}_c)$. If we assume that there is no correlation between $\dot{\varepsilon}_x$ and $\dot{\varepsilon}_p$, we obtain the linewidth of the scattering photon:

$$v_s = 2v_p + \frac{v_p}{(1+\Gamma/\kappa)^2}. \tag{S70}$$

Based on Eq. (S70), we find that in the reversed regime, the linewidth of the corresponding Stokes light is approximately equal to twice the linewidth of the pump when ignoring the thermal noise, which is also suggested by our numerical simulation. Then, according to the similar method used in the normal regime ($\Gamma \ll \kappa$), the contribution of thermal noise can be calculated separately, which leads to the full expression of the linewidth of the phonon and scattering photon.

$$v_s = 2v_p + v_p/(1+\Gamma/\kappa)^2 + v_{mT} \approx 2v_p + v_{mT}, \tag{S71}$$

$$v_m = \frac{v_p}{(1+\kappa/\Gamma)^2} + v_{mT}, \text{ (phonon linewidth formula is the same in both regimes)}$$

$$v_{mT} = \frac{\Gamma n_{th}}{2n}. \qquad \text{(ST linewidth formula is the same in both regimes)}$$

In summary, as the intensity of the scattering light is much higher than the intensity of the pump light in the reversed regime (see Fig. S2), ignoring the high-order terms of the scattering light, which is equivalent to ignoring the randomness of $\tilde{\Delta}$, will lead to the deficient theory of linewidth analysis in the reversed regime. An incorrect scattering photon linewidth will be obtained in both theoretical analysis and numerical simulation[3]. We first showed that in the reversed regime of the optomechanical system ($\Gamma \gg \kappa$), the linewidth of the scattering photon is no longer as narrow as that of the stimulated Brillouin system owing to the jitter in the equilibrium position. The derived analytical expression suggests that the linewidth of the scattering photon in the reversed regime will always be more than twice the pump linewidth.

## S4. Detected spectrum

The spectrum around the oscillation frequency detected by the high-speed photoreceiver will represent the mechanical motion, even though the light signal is detected.

Assume that the mechanical motion is $x = x_1 e^{i\Omega_m t} + x_1^* e^{-i\Omega_m t}$. Then, the power spectrum of the mechanical motion around the oscillation frequency is proportional to that of $x_1$. Assume that the optical amplitude in the cavity has the form of $a = a_0 + a_1 e^{i\Omega_m t} + a_{-1} e^{-i\Omega_m t}$. As the photoreceiver detects the intensity of the output light, which is proportional to $|s_{in} - \sqrt{\kappa_{ex}} a|^2$, the signal detected by the photoreceiver around the oscillation frequency is proportional to $(a_0 a_{-1}^* + a_0^* a_1) e^{i\omega t} + (a_0^* a_{-1} + a_0 a_1^*) e^{-i\omega t}$. Consequently, the detected power spectrum around the oscillation frequency is proportional to that of $a_0 a_{-1}^* + a_0^* a_1$.

We can substitute $x$ and $a$ into Eq. (1) in the main text. Then,

$$a_1 = \frac{-iGx_1 a_0}{\frac{\kappa}{2} + i(\Delta - \omega)}, \tag{S72}$$

$$a_{-1} = \frac{-iGx_1^* a_0}{\frac{\kappa}{2} + i(\Delta + \omega)}, \tag{S73}$$

where $G = g_0/x_{zpf}$ is the optical frequency shift per displacement. Hence, the detected power spectrum around the oscillation frequency is proportional to

$$a_0 a_{-1}^* + a_0^* a_1 = \frac{\kappa_{ex}}{(\frac{\kappa^2}{4} + \Delta^2)} \left( \frac{-iG}{\frac{\kappa}{2} + i(\Delta - \omega)} + \frac{iG}{\frac{\kappa}{2} - i(\Delta + \omega)} \right) x_1 |s_{in}|^2, \tag{S74}$$

which eliminates the phase fluctuation in the pump and only reflects that of the mechanical motion.

# S5. Measurement setup

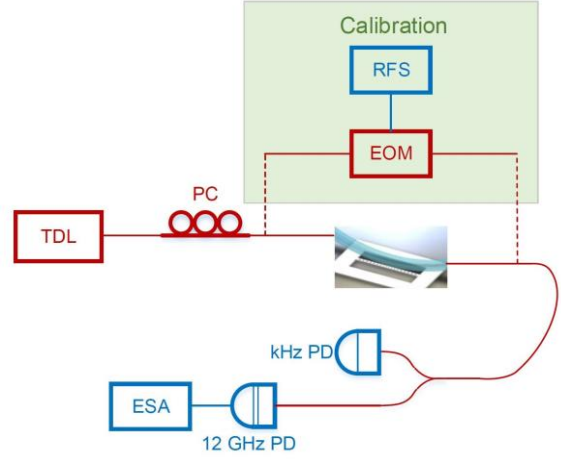

**Figure S3.** Measurement setup. TDL: tunable diode laser; PC: polarization controller; kHz PD: low-speed photodetector; 12 GHz PD: high-speed photoreceiver; ESA: electronic spectrum analyser; RFS: radio frequency (RF) signal source; EOM: intensity electro-optic modulator. The red connections represent the optical signals and the blue connections represent the electrical signals. The RFS and EOM are used to calibrate the intensity of the oscillation signal.

# S6. Measuring optomechanical coupling rate ($g_0$)

We measure the optomechanical coupling rate using a method similar to that proposed by M. L. Gorodetksy *et al.* and K. Balram *et al.*[12,13] and Z. Huang *et al.*[14] The power of the optical signal detected by the photoreceiver around the oscillation frequency is related to the mechanical vibration intensity with $g_0$. This power can be measured quantitatively by the electronic spectrum analyser with the assistance of the intensity electro-optic modulator for calibration. Thus, the optomechanical coupling rate ($g_0$) can be obtained.

Assume that the mechanical vibration is $x(t) = x_0 \cos(\Omega_m t)$, substitute it into Eq. (1) in the main text, and assume a stable optical input. Then, the normalized optical amplitude in the cavity can be expressed as

$$a(t) = s_{in}\sqrt{\eta\kappa}\mathcal{L}(0)\left(1 - \frac{ix_0 g_0 \mathcal{L}(\Omega_m)}{2x_{zpf}}e^{-i\Omega_m t} - \frac{ix_0 g_0 \mathcal{L}(\Omega_m)}{2x_{zpf}}e^{i\Omega_m t}\right), \tag{S75}$$

where function $\mathcal{L}(\Omega)$ is defined as

$$\mathcal{L}(\Omega) = \frac{1}{-i(\Delta+\Omega)+\kappa/2} \tag{S76}$$

to simplify the derivation. The normalized output optical amplitude is

$$s_{out}(t) = s_{in} - \sqrt{\kappa_{ex}}\,a(t) \tag{S77}$$

and the relation between the output optical power ($I_{OM}$) and the normalized output optical amplitude is

$$I_{OM}(t) = \hbar\omega_o |s_{out}(t)|^2. \tag{S78}$$

The output optical power affected by the optomechanical cavity system has the form

$$I_{OM}(t) = I_{0OM} + I_{1OM}\cos(\Omega_m t + \theta). \tag{S79}$$

The power detected by the electrical spectrum analyser around the oscillation frequency should be proportional to $I_{1OM}^2$, i.e. $P_{ESA,OM} = k_{ESA}I_{1OM}^2$. Hence, we focus on the term $I_{1OM}^2$. From Eqs. (S78) and (S79), we have

$$I_{1OM}^2 = \frac{1024\Delta^2\eta^2\kappa^2\left((-1+\eta)^2\kappa^2+\Omega_m^2\right)}{(4\Delta^2+\kappa^2)^2\left((4\Delta^2+\kappa^2)^2+8\Omega_m^2(-4\Delta^2+\kappa^2+2\Omega_m^2)\right)}\left(\frac{\hbar\omega_L x_0 g_0}{x_{zpf}}\right)^2|s_{in}|^4, \tag{S80}$$

where $\eta = \kappa_{ex}/\kappa$. The value of Eq. (S80) is related with optical detuning ($\Delta$). When $\Delta = \frac{\sqrt{\kappa^2+4\Omega_m^2}}{2\sqrt{3}}$, its value will be maximum as

$$I_{1OM}^2 = \frac{27\eta^2\kappa^2\left((-1+\eta)^2\kappa^2+\Omega_m^2\right)}{(\kappa^2+\Omega_m^2)^3}\left(\frac{\hbar\omega_L x_0 g_0}{x_{zpf}}\right)^2|s_{in}|^4. \tag{S81}$$

During the experimental measurement, we fine-tuned the wavelength of the input laser to obtain the maximum signal in the electrical spectrum analyser. Thus, Eq. (S81) can be used to estimate the power detected by the electrical spectrum analyser around the oscillation frequency.

The ration between $x_0$ and $x_{zpf}$ can be obtained by $x_0^2/x_{zpf}^2 = 4n$, where $n$ is the number of phonons in the cavity and it can be obtained by the measurement shown in Fig. 4(a) of the main text.

To calibrate the power obtained by the electrical spectrum analyser, we used an intensity electro-optic modulator, the procedure for which is the same as that performed by Z. Huang et al.[14] By determining the power detected by the electrical spectrum analyser around the oscillation frequency, we can get the optomechanical coupling rate ($g_0/2\pi$) from Eq. (S81), which is 1.9 MHz for the hetero optomechanical crystal cavity.

## S7. Coupled optical power

As cavities are optically coupled by a tapered fibre in our experiment, only a small portion of the light is coupled to our cavity. The expression for estimating the coupled optical power is

$$P_{coupled} = \hbar\omega_L\kappa|a|^2 = 4\frac{\kappa_{ex}\kappa}{\kappa^2+4\Delta^2}P_{in}. \tag{S82}$$

which is the same as that used by W. Jiang et al.[15] In this equation, $P_{in} = \hbar\omega_L|s_{in}|^2$ is the output power of the laser source.

## S8. The power of scattering light

The power of the scattering light shown in Fig. 4(a) of the main text is inferred from the phonon number ($n_m$), which can be obtained by the signal detected by the electronic spectrum analyser. From Eq. (S77), the output power of the scattering light (with frequency $\omega_L$-$\Omega_m$) can be expressed as

$$P_{1out} = \frac{4\kappa_{ex}^2 g_0^2}{\left(4\Delta^2 + \kappa^2\right)\left(\kappa^2 + 4(\Delta - \Omega_m)^2\right)} 4n_m P_{in}. \tag{S83}$$

## S9. Photonic and phononic band of the structure

The structure of the optomechanical crystal with an acoustic radiation shield is shown in Fig. S4a. The unit cell of the nanobeam periodic structure is shown in Fig. S4b. The height ($h$) of the structure is 220 nm and the width ($w$) of the structure is 460 nm. The period ($d$) and the radius of the air hole ($r$) are 493 nm and 136 nm, respectively, for the nanobeam periodic structure. The photonic and phononic bands of the nanobeam periodic structure are plotted in Fig. S4c and d, respectively. The frequencies of the optical and mechanical modes are indicated with red dashed lines. The unit cell of the acoustic radiation shield is shown in Fig. S4e. The phononic bands are plotted in Fig. S4f. The optical mode is confined by the nanobeam periodic structure as shown in Fig. S4c. However, the high-frequency mechanical mode exceeds the mechanical bandgap of the nanobeam periodic structure as shown in Fig. S4d. Consequently, the mechanical mode is confined by the acoustic radiation shield as shown in Fig. S4f.

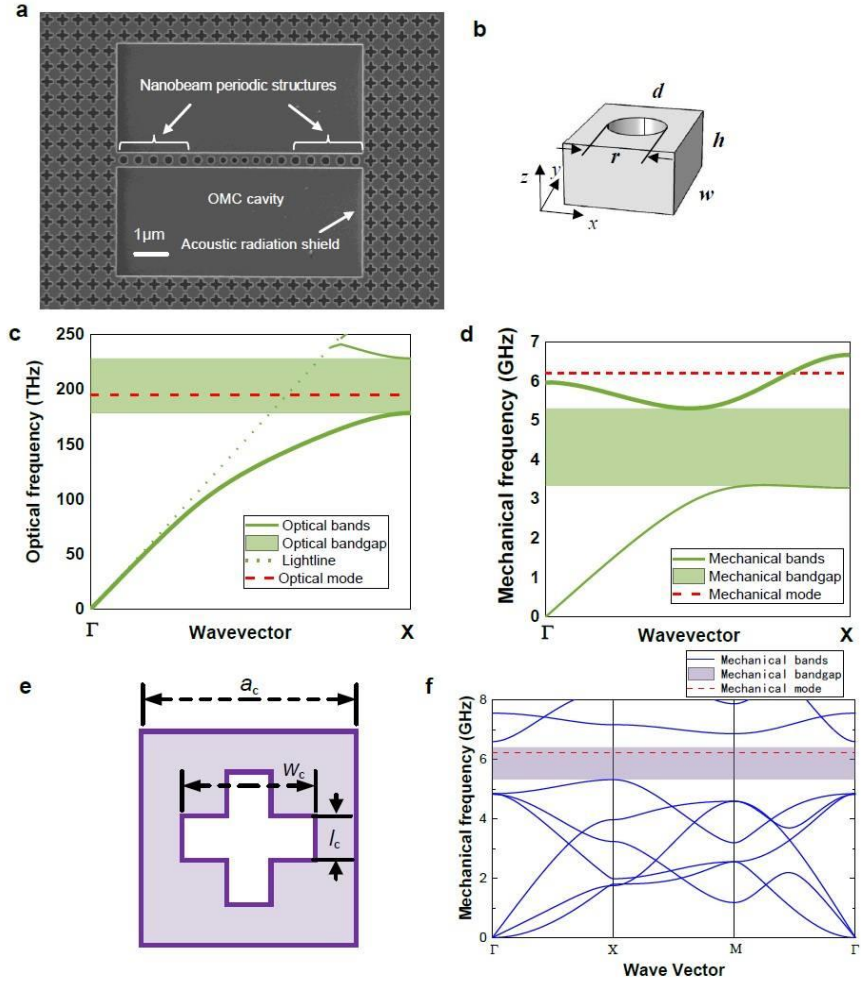

**Figure S4.** (**a**) Structure of optomechanical crystal cavity with acoustic radiation shield. (**b**) Schematic of the unit cell of the nanobeam structure. The (**c**) photonic and (**d**) phononic band of the nanobeam periodic structure. (**e**) Schematic of the unit cell of the acoustic radiation shield. (**f**) Phononic band of the acoustic radiation shield.

## S10. Phonon Schawlow–Townes linewidth

From Eq. (S52), we can get the phonon linewidth formula, which is similar to the Schawlow–Townes limit of lasers. The linewidth formula, $v_{mT} = \Gamma n_{th}/(2n)$, indicates that the phonon linewidth owing to the thermal noise is inversely proportional to the excited intra-cavity number of phonons. Thus, when the pump power reaches the threshold, the large accumulation of phonon energy will lead to considerable narrowing of the phonon linewidth.

In Fig. 4c of the main text, we use this formula to fit the linewidth curve in the laser regime. As the 1D nanobeam micro-cavity used in our experiments has an extremely small optical mode volume, the thermal effect of the micro-cavity must be considered. However, it is difficult to measure the exact temperature of the cavity. Hence, $n_{th}$ (proportional to $T$) is set to be a fitting parameter. The power spectra obtained from our experiments are used to calculate the intra-cavity number of phonons. By using the relationship[1] $n_{th} \approx k_B T/(\hbar\Omega_m)$, after we get $n_{th}$, we can also derive the cavity temperature $T$. The fitted temperature $T$ is around 600 K. Note that as the temperature of the micro-cavity is related to the injected pump power

(thus changing with the injected power), the fitted value can reflect only the average temperature level when the pump power is increased.